\documentclass[twocolumn,aps,prl,superscriptaddress,floatfix,noeprint]{revtex4-1}

\usepackage{graphicx,tabularx}
\usepackage{amsmath,amsfonts,amssymb}
\usepackage{bm,dsfont}
\usepackage[colorlinks=true, citecolor=blue, linkcolor=red, urlcolor=blue]{hyperref}
\usepackage[all]{hypcap}
\usepackage{xcolor}
\usepackage[caption=false]{subfig}

\graphicspath{{figures/}}

\def\blankpage{%
      \clearpage%
      \thispagestyle{empty}%
      \addtocounter{page}{-1}%
      \null%
      \clearpage}

\DeclareMathOperator{\diag}{diag}
\DeclareMathOperator{\sgn}{sgn}

\begin{document}

\title{Aharonov-Bohm Oscillations in Minimally Twisted Bilayer Graphene}
\author{C. De Beule}
%\email{c.de-beule@tu-braunschweig.de}
\affiliation{Institute for Mathematical Physics, TU Braunschweig, 38106 Braunschweig, Germany}
\author{F. Dominguez}
%\email{f.dominguez@tu-braunschweig.de}
\affiliation{Institute for Mathematical Physics, TU Braunschweig, 38106 Braunschweig, Germany}
\author{P. Recher}
%\email{p.recher@tu-braunschweig.de}
\affiliation{Institute for Mathematical Physics, TU Braunschweig, 38106 Braunschweig, Germany}
\affiliation{Laboratory for Emerging Nanometrology, 38106 Braunschweig, Germany}
\date{\today}

\begin{abstract}
We investigate transport in the network of valley Hall states that emerges in minimally twisted bilayer graphene under interlayer bias. To this aim, we construct a scattering theory that captures the network physics. In the absence of forward scattering, symmetries constrain the network model to a single parameter that interpolates between one-dimensional chiral zigzag modes and pseudo-Landau levels. Moreover, we show how the coupling of zigzag modes affects magnetotransport. In particular, we find that scattering between parallel zigzag channels gives rise to Aharonov-Bohm oscillations that are robust against temperature, while coupling between zigzag modes propagating in different directions leads to Shubnikov-de Haas oscillations that are smeared out at finite temperature.
\end{abstract}

\maketitle

Twisted bilayer graphene (TBG) consists of two graphene layers stacked with a relative twist, leading to a moir\'e pattern of alternating stacking domains that drastically alters the electronic structure \cite{LopesDosSantos2007,SuarezMorell2010,Bistritzer2010,Li2010}. Over the past few years, TBG attracted great interest due to the discovery of exotic phenomena in magic-angle TBG \cite{Kim2017a,Cao2018a,Cao2018,Yankowitz2019,Sharpe2019,Kerelsky2019,Choi2019,Cao2019}. At minimal twist angles $\theta\sim0.1^\circ$, TBG also exhibits interesting physics. In this case, the lattice relaxes into sharply defined triangular $AB$/$BA$ stacking domains \cite{Nam2017,Yoo2019,Walet2019}. When a potential bias $U$ is applied between the layers, a local gap is opened in the $AB$/$BA$ stacking regions with valley Chern number $N_K = -N_{K'} \approx \pm \sgn(U/\gamma_\perp)$ where $\pm$ corresponds to $AB$ or $BA$ stacking respectively, and $\gamma_\perp$ is the interlayer hopping \cite{Martin2008,Zhang2013}. Consequently, each valley and spin hosts two chiral modes along $AB$/$BA$ domain walls that propagate in opposite directions for different valleys \cite{Zhang2013,Vaezi2013,Ju2015,Yin2016}. When the Fermi energy is tuned in the gap, the low-energy excitations are therefore entirely due to a triangular network of valley Hall states \cite{San-jose2013,Efimkin2018,Ramires2018,Huang2018,Sunku2018}. Recently, it was observed from microscopic calculations that the network gives rise to one-dimensional (1D) chiral zigzag (ZZ) modes \cite{Fleischmann2020,Tsim2020}. However, current network theories \cite{Efimkin2018} cannot reproduce these results and recent transport experiments that reported interference oscillations are incompatible with decoupled 1D chiral modes \cite{Rickhaus2018,Xu2019}. In particular, Aharonov-Bohm (A-B) oscillations were observed in the presence of a magnetic field perpendicular to the layers \cite{Xu2019}. In contrast to other topological systems where A-B oscillations arise due to interference between 1D topological channels at the boundaries \cite{Ji2003,Cho2015,MacIejko2010,Virtanen2011,Dolcini2011}, the network in TBG extends over the whole system. To our knowledge, a transport theory for these observations is lacking, most likely due to the large computational cost of standard methods at such small twist angles since the number of carbon atoms per moir\'e cell is of the order of $10^4 \left( \theta^\circ \right)^{-2}$. 
\begin{figure}
\centering
\includegraphics[width=\linewidth]{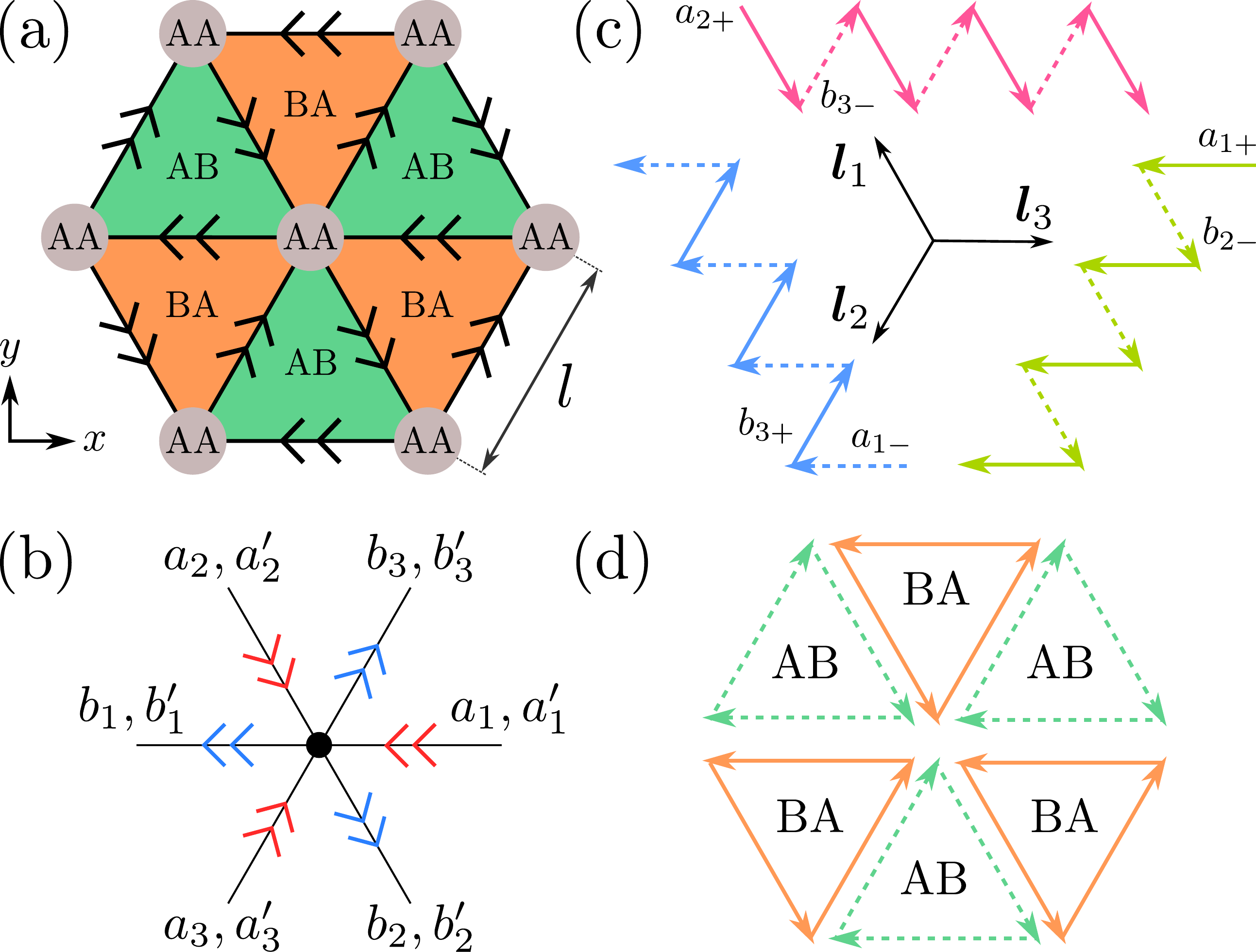}
\caption{(a) Stacking domains of TBG showing the network of valley Hall states (for a single valley and spin) that emerges upon applying an interlayer bias. $AB$/$BA$ domain walls and $AA$ regions correspond to the links and scattering nodes of the network, respectively. (b) Unit cell of the network. (c) Triplet of 1D chiral ZZ modes ($\phi = 0$) along directions $\bm l_j$ ($j=1,2,3$) and (d) pseudo-Landau levels ($\phi = \pi/2$). Solid (dashed) arrows correspond to antisymmetric (symmetric) superpositions of valley Hall states along the same link.}
\label{fig:fig1}
\end{figure}

In this Letter, we construct a phenomenological scattering theory for the chiral network that emerges in TBG under interlayer bias and investigate magnetotransport through the network. The network consists of nodes characterized by an $S$ matrix, and links between nodes along which modes propagate freely \cite{Chalker1988}. In this case, the links are given by $AB$/$BA$ domain walls supporting two chiral channels per valley and spin, and the nodes correspond to metallic $AA$ regions, as illustrated in Fig.\ \ref{fig:fig1}(a). In contrast to previous network models \cite{Efimkin2018}, we take into account scattering at the nodes between the two channels, which is crucial to obtain agreement with microscopic calculations and experiments. While the two valley Hall states do not mix along links in the absence of disorder, it is not \emph{a priori} clear why they should remain decoupled as they reach the $AA$ regions, where the local gap induced by the interlayer bias vanishes. 

In the absence of forward scattering, we find that the network physics is controlled by the phase shift $\phi$ acquired after $120^\circ$ deflections, which tunes between 1D chiral ZZ modes and pseudo-Landau levels \cite{Ramires2018}. We then include forward scattering and discuss the different scattering mechanisms between ZZ modes. Depending on the coupling of ZZ modes, we find two distinct transport regimes: a quasi-1D regime for which only parallel ZZ channels are coupled, and a 2D regime where ZZ modes propagating in different directions are coupled. In the quasi-1D regime, the magnetoconductance displays A-B oscillations, while in the 2D regime they are accompanied by Shubnikov-de Haas (SdH) oscillations. Moreover, at finite temperature, we find that the A-B oscillations are robust, while the SdH oscillations are smeared out.

\emph{Network model} --- We consider a network with two chiral modes along each link which scatter at nodes that form a triangular lattice, as illustrated in Fig.\ \ref{fig:fig1}(a). Each scattering node has six incoming and six outgoing modes as shown in Fig.\ \ref{fig:fig1}(b). We label the nodes by their position vector $\bm r_{ij} = i \bm l_1 + j \bm l_2$ where $\bm l_{1,2} = l (-1/2,\pm \sqrt{3}/2)$ are moir\'e lattice vectors with $l = a/2\sin(\theta/2)$ the moir\'e lattice constant and where $a$ is the lattice constant of graphene. Incoming modes are denoted as $a_{ij} = (a_{1,ij},a_{2,ij},a_{3,ij})$ and $a_{ij}'$ for the two chiral channels, while outgoing modes are denoted as $b_{ij}$ and $b_{ij}'$, such that $(b_{ij},b_{ij}')^t=\mathcal S(a_{ij},a_{ij}')^t$ with $\mathcal S$ the $S$ matrix relating incoming to outgoing modes. We do not consider intervalley scattering as the moir\'e pattern varies slowly on the interatomic scale for small twists. 

To constrain the $S$ matrix, we take into account the symmetries of TBG under interlayer bias. At small twist angles, the symmetries of TBG become independent of the twist center \cite{Po2018,Zou2018} so that we do not have to consider a specific lattice realization. Symmetries that preserve the valley are given by $C_3$ and $C_2 T$, where $T$ is (spinless) time-reversal symmetry and $C_3$ and $C_2$ are rotations by $2\pi/3$ and $\pi$ about the $z$-axis with respect to the center of an $AA$ region, respectively. Note that $C_2$ exchanges both the A and B sublattices and valleys. In the basis of Fig.\ \ref{fig:fig1}(b), they impose the following conditions \footnote{See supplemental material [url to be added].}:
\begin{equation}
C_3: \quad  \mathcal S = C_3 \mathcal S C_3^{-1}, \qquad C_2 T: \quad  \mathcal S = \mathcal S^t,
\end{equation}
where $C_3$ corresponds to a cyclic permutation of the incoming modes $(a_1,a_1') \rightarrow (a_2,a_2') \rightarrow (a_3,a_3') \rightarrow (a_1,a_1')$ and similar for outgoing modes.

To proceed, we first neglect forward scattering, which is a good starting point as the wave-function overlap between incoming and outgoing modes is larger for deflections than for forward scattering, due to the network geometry \cite{Qiao2014a}. Note that this is impossible if the two valley Hall states are decoupled, as in this case there is a lower bound on forward scattering \cite{Efimkin2018}. We find that the $S$ matrix is given up to a unitary transformation by
\begin{equation} \label{eq:S0}
\mathcal S = \frac{1}{2} \begin{pmatrix}
S_{\phi,\phi} & S_{0,\pi} \\
S_{\pi,0} & -S_{-\phi,-\phi}
\end{pmatrix},
\end{equation}
with
\begin{equation}
S_{\vartheta,\psi} =
\begin{pmatrix}
0 & e^{i\vartheta} & e^{i\psi} \\
e^{i\psi} & 0 & e^{i\vartheta} \\
e^{i\vartheta} & e^{i\psi} & 0
\end{pmatrix},
\end{equation}
and where $\phi \in [0,\pi/2]$ \cite{Note1}. The incoming modes of a given node are also related to outgoing modes of neighboring nodes by $a_{ij}= \lambda (b_{1,i-1j-1},b_{2,i+1j},b_{3,ij+1})$ where $\lambda=\exp(iEl/\hbar v)$ is the dynamical phase picked up along a link with $v$ the velocity of the modes, which we assume is equal for the two valley Hall states. Using Bloch's theorem, we find $(a, a')^t = \lambda \left[ \mathds 1_2 \otimes \mathcal M(\bm k) \right] \mathcal S (a,a')^t$ where $\mathcal M(\bm k) = \textrm{diag} \left( e^{ik_3}, e^{ik_1}, e^{ik_2} \right)$ with $k_j = \bm k \cdot \bm l_j$ ($j=1,2,3$) and $\bm l_3 = -(\bm l_1+\bm l_2)$. For convenience, we suppressed the momentum index on the amplitudes $a$ and $a'$. The network energy bands are then given by the phase of the eigenvalues of $\left[ \mathds 1_2 \otimes \mathcal M(\bm k) \right] \mathcal S$ \cite{Efimkin2018,Pal2019}.
\begin{figure}
\centering
\includegraphics[width=\linewidth]{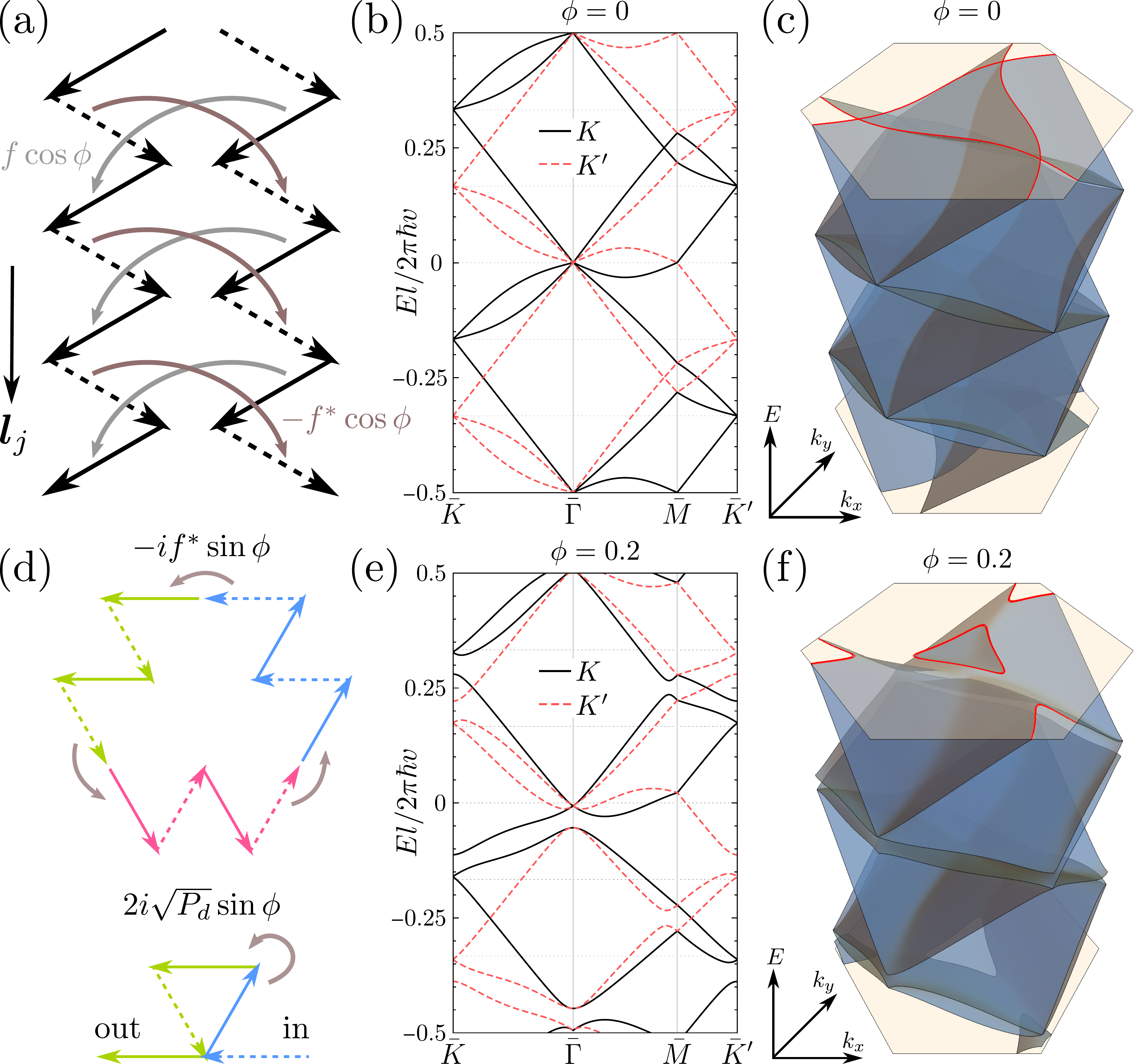}
\caption{(a) Coupling between parallel ZZ channels due to forward scattering. (b) Network bands for $\phi=0$, $P_{f1} = 0$, and $P_{f2} = 0.04$ along high-symmetry lines, where solid (dashed) lines correspond to $K$ ($K'$) and (c) in the network Brillouin zone for $K$ where the Fermi surface is shown in red. (d) Couplings between ZZ modes that propagate in different directions. (e,f) Spectrum for $\phi=0.2$ and $P_{f1} = P_{f2} = 0.02$.}
\label{fig:fig2}
\end{figure}

Hence, in the absence of forward scattering the network physics is tuned only by the intrachannel deflection phase shift $\phi$ appearing in Eq.\ \eqref{eq:S0}, which we treat as a phenomenological parameter. For $\phi = 0$, we find that the network spectrum is given by
\begin{equation} \label{eq:spectrum1}
E_{j,n}(\bm k) = \frac{\hbar v}{2l} \left( 2 \pi n + k_j \right),
\end{equation}
where $n\in \mathbb Z$. Since the energy enters in the dynamical phase, the network spectrum is periodic in energy, in this case with period $\pi\hbar v/l$. To understand this result, we perform a unitary transformation $U = e^{-i\pi \sigma_y/4} e^{i\phi\sigma_z/2} \otimes \mathds 1_3$ on $\mathcal S$, which transforms the original basis $a,a'$ to a basis of symmetric and antisymmetric superpositions of valley Hall states on the same link: $a_\pm = ( a e^{i \phi / 2} \mp a' e^{-i \phi / 2})/\sqrt{2}$ and similar for outgoing modes. For $\phi=0$, we find that there are only interchannel deflections in the new basis that proceed counterclockwise (clockwise) for $a_+$ ($a_-$), see Fig.\ \ref{fig:fig1}(c) \cite{Note1}. This gives rise to three independent chiral ZZ channels with velocity $v/2$ due to the ZZ motion. The opposite limit, $\phi = \pi/2$, results in three doubly-degenerate flatbands per period $2\pi\hbar v/l$, that we identify with pseudo-Landau levels. Now there are only intrachannel deflections where $a_+$ ($a_-$) modes perform counterclockwise (clockwise) orbits around $BA$ ($AB$) domains, see Fig.\ \ref{fig:fig1}(d). Hence the network modes are localized, leading to flatbands. 

\emph{Coupling of zigzag modes ---}
We now discuss the couplings between ZZ modes. To this end, we include intra- and interchannel forward scattering with probabilities $P_{f1}$ and $P_{f2}$, respectively, but take the same probability $P_d$ for intra- and interchannel deflections. Different deflection probabilities lead to additional couplings, but these do not change our results qualitatively. In this case, current conservation gives $4P_d +  P_f= 1$ with $P_f=P_{f1} + P_{f2}$, and the $S$ matrix in the new basis becomes
\begin{equation} \label{eq:S1}
U \mathcal S U^\dag = \begin{pmatrix} 
S_1 & S_2^\dag \\
S_2 & -S_1^\dag
\end{pmatrix},
\end{equation}
where $S_1 = 2 i \sqrt{P_d} \sin \phi \, s_0 + f \cos \phi \, \mathds 1_3$, $S_2 = 2 \sqrt{P_d} \cos \phi \, s_0 + i f \sin \phi \, \mathds 1_3$, with $f = \sqrt{P_{f2}} + i \sqrt{P_{f1}}$ and $[s_0]_{nm} = \delta_{n1} \delta_{3m} + \delta_{n2} \delta_{1m} + \delta_{n3} \delta_{2m}$. Here, $S_1$ ($S_2$) corresponds to intrachannel (interchannel) processes in the new basis, $a_+ \rightarrow b_+ (b_-)$.

When $\phi=0$, only parallel ZZ channels are coupled by $S_1$ in Eq.\ \eqref{eq:S1} via intrachannel forward scattering for $a_\pm$ modes with probability $|f|^2=P_f$, as shown in Fig.\ \ref{fig:fig2}(a). Hence, the network effectively decouples into three independent quasi-1D systems. In general, however, ZZ channels propagating in different directions are coupled through two processes, see Fig.\ \ref{fig:fig2}(d). One process is due to interchannel forward scattering for $a_\pm$ modes with probability $P_f \sin^2 \phi$, while the other is due to counterclockwise (clockwise) intrachannel deflections for $a_+$ ($a_-$) with probability $4P_d \sin^2 \phi$. Both processes lead to anti-crossings in the network spectrum as shown in Figs.\ \ref{fig:fig2}(e) and (f). In this case, the network corresponds to a true two-dimensional (2D) percolating system.
\begin{figure}
\centering
\includegraphics[width=\linewidth]{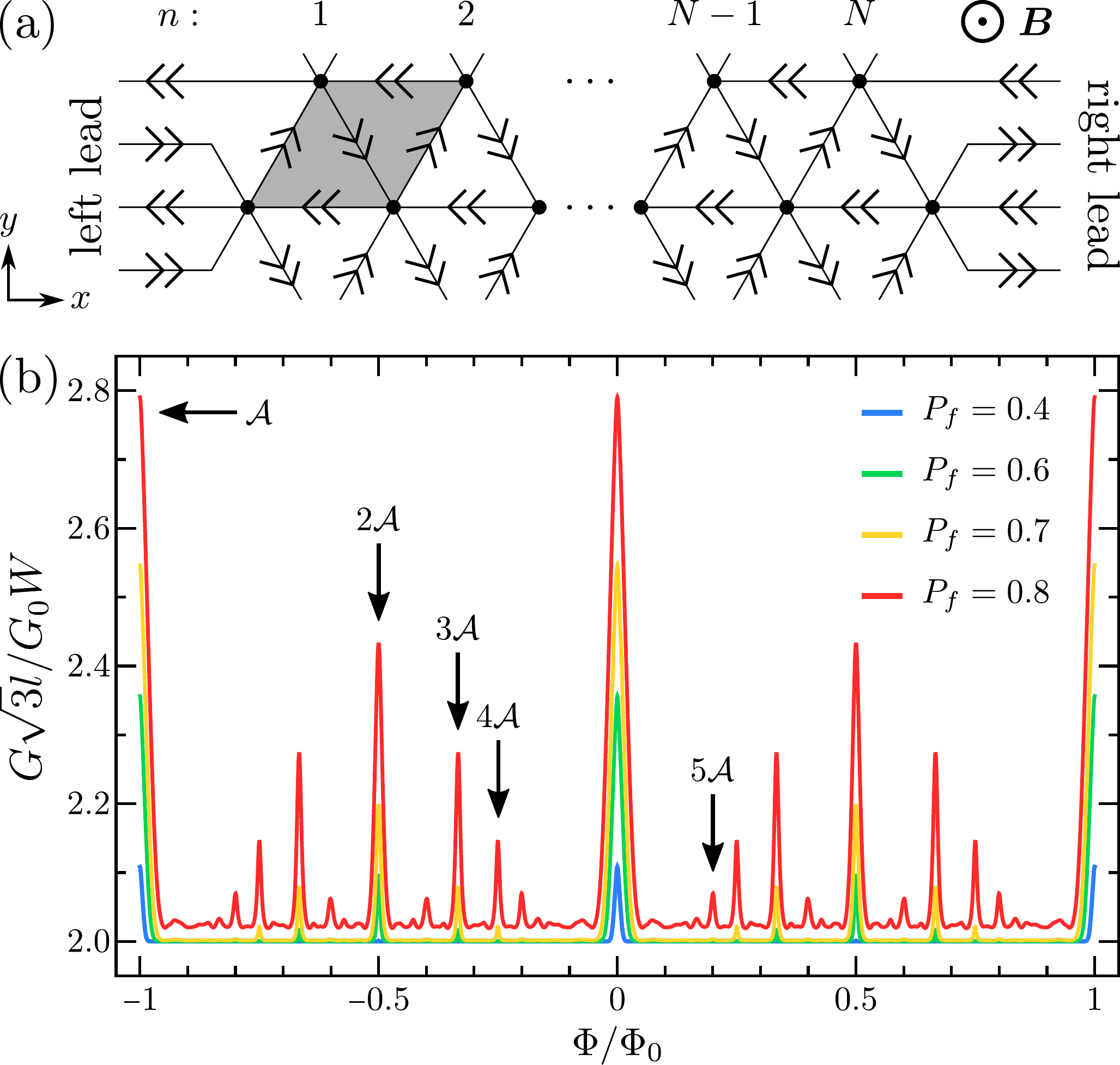}
\caption{(a) Unit cell of a network strip with length $L=Nl$ where the moir\'e cell with area $\mathcal A=\sqrt{3} \, l^2/2$ is given by the gray diamond. (b) Magnetoconductance for $L = 10 \, l$ and width $W\gg L$ in the quasi-1D regime ($\phi=0$) for several $P_f$ as a function of $\Phi=B\mathcal A$. Arrows indicate the encircled area of interfering paths corresponding to A-B resonances.}
\label{fig:fig3}
\end{figure}

\emph{Magnetotransport} ---  Given the effective dimensionality of the two coupling regimes, one expects a different behavior in the presence of a magnetic field $\bm B = B \bm e_z$ perpendicular to the network. If we assume the magnetic field is sufficiently small such that the structure of the $S$ matrix is unchanged, it enters only though the Peierls phase accumulated during propagation along links. In the Landau gauge $\bm A = Bx \bm e_y$, the Peierls phase along a link starting at $x=ml/2$ ($m \in \mathbb Z$) is zero for horizontal links and for an upward or downward diagonal link, $\Phi_{\pm}(m) = \mp \pi \left( m + 1/2 \right) \Phi/\Phi_0$, respectively. Here, $\Phi_0 = h/e$ is the flux quantum and $\Phi = B \mathcal A$ is the flux through a moir\'e cell, comprising an $AB$ and $BA$ triangle with $\mathcal A = \sqrt{3} \, l^2/2$. To investigate magnetotransport, we calculate the zero-bias differential conductance for a network strip of length $L$ and width $W\gg L$, whose unit cell is shown in Fig.\ \ref{fig:fig3}(a), 
\begin{equation}
\frac{G}{G_0} = N_i \mathcal T, \qquad N_i = \frac{4W}{\sqrt{3}l},
\end{equation}
with $G_0=4e^2/h$ and $N_i$ the number of channels, where the transmission probability $\mathcal T$ is calculated with the transfer matrix that connects the modes at the right and left leads \cite{Chalker1988,Note1}. Note that $\mathcal T$ is the same for both valleys in the limit $W\gg L$.

\begin{figure}
\centering
\includegraphics[width=\linewidth]{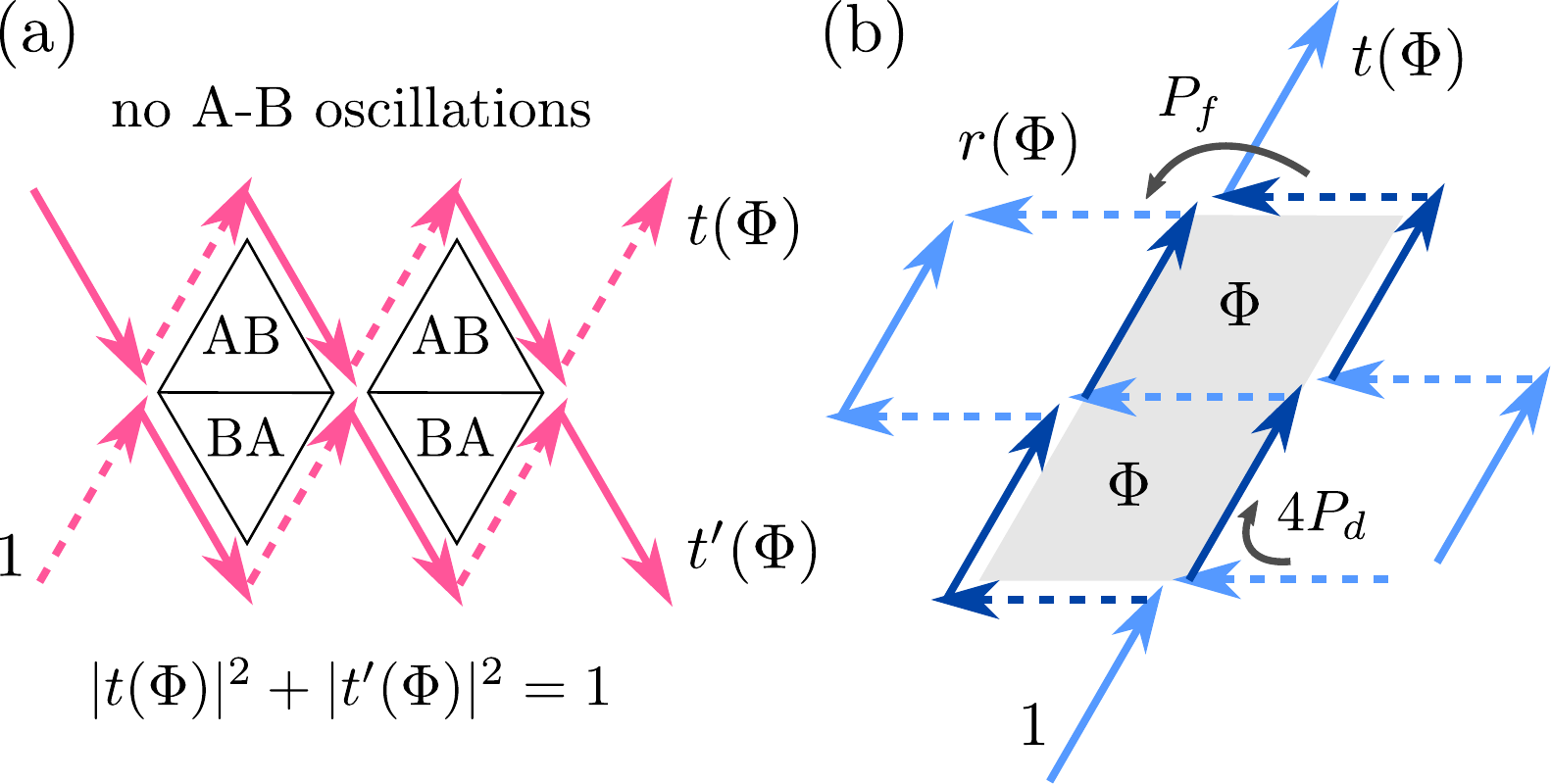}
\caption{Processes giving rise to the background (a) and A-B oscillations (b) in the quasi-1D regime [see Fig.\ \ref{fig:fig3}(b)], where $4P_d$ and $P_f$ are the probabilities to stay in the same ZZ branch or jump to a neighboring parallel branch, respectively. Dark blue lines in (b) show two paths enclosing a flux $2\Phi$, and $t$, $t'$, and $r$ are examples of transmission and reflection amplitudes.}
\label{fig:fig4}
\end{figure}

\begin{figure}
\centering
\includegraphics[width=\linewidth]{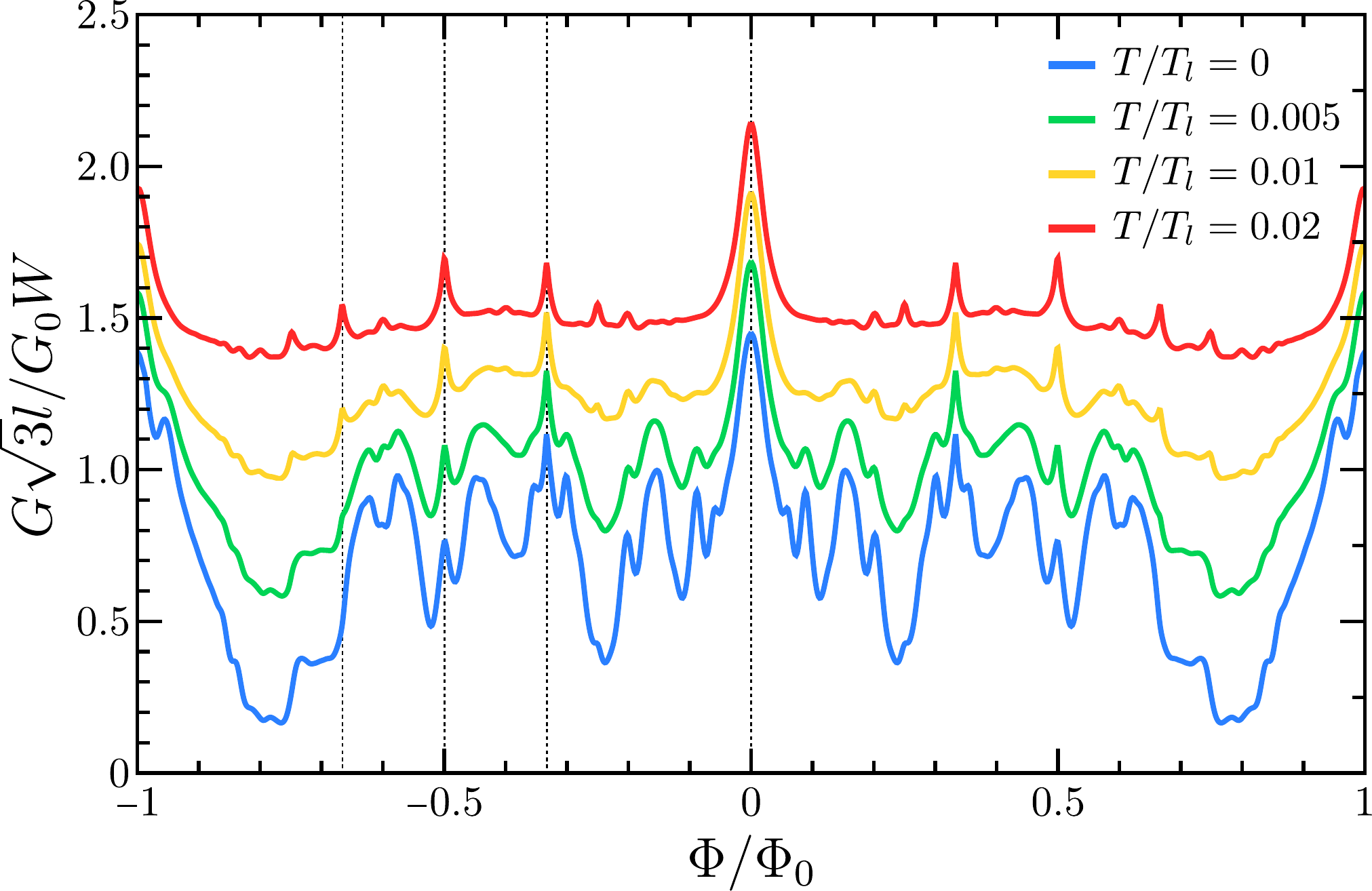} 
\caption{Temperature dependence of the magnetoconductance for a network strip of length $L=10 \, l$ and width $W\gg L$ with $\phi=0.2$, $P_{f1,2}=0.3$, and $E_F=\pi\hbar v/6l$ where $T_l = 2\pi \hbar v/k_Bl \approx 3400 \, (\theta^\circ)$~K. For visibility, the curves are shifted downwards for $T/T_l = 0$, $0.005$, and $0.01$.}
\label{fig:fig5}
\end{figure}

The magnetoconductance in the quasi-1D regime is shown in Fig.\ \ref{fig:fig3}(b). We find that $G(\Phi)$ depends only on $P_f=P_{f1}+P_{f2}$ and observe periodic resonances on a constant background. This plateau is due to ZZ modes that are chiral in the transport direction ($\bm l_3$). Hence, they are always fully transmitted and cannot contribute to A-B oscillations, see Fig.\ \ref{fig:fig4}(a). On the other hand, when forward scattering is present, parallel ZZ channels are coupled so that ZZ modes propagating oppositely to the transport direction ($\bm l_{1,2}$) can also contribute, giving rise to A-B oscillations, see Fig.\ \ref{fig:fig4}(b). By including the contribution of the $\bm l_3$ branch, and interfering paths of lengths $2L$ and $2L+3l$ from the $\bm l_{1,2}$ branches, we find
\begin{equation} \label{eq:Ga2}
\frac{G}{G_0} \simeq \left( 2 + 2 P_f^{2L/l+1} \left[ 1 + P_f (4P_d)^2 F_L(\Phi)^2 \right] \right) \frac{W}{\sqrt{3}l},
\end{equation}
where $F_L(\Phi) = \sin \left[ \pi (2L/l-1) \Phi/\Phi_0 \right]/\sin (\pi \Phi/\Phi_0)$, reproducing the main resonance and background, where the peak width $\Delta \Phi = 2\Phi_0/(2L/l-1)$. The main A-B period corresponds to one flux quantum per moir\'e cell as paths encircling a single $AB$ or $BA$ triangle do not occur in the quasi-1D regime. In general, A-B resonances occur at $\Phi = \Phi_0/n\mathcal A$ with $n$ a nonzero integer, due to interfering paths encircling an area equal to $n$ times the moir\'e cell. Note that the conductance is energy-independent in this case, as interfering paths always have the same length in the quasi-1D regime, and therefore pick up the same dynamical phase. This implies that the A-B resonances are robust against temperature \cite{Virtanen2011}. We find that the A-B oscillations persist for finite $\phi$ (2D regime), but are accompanied by SdH oscillations resulting from network Landau levels, as shown in Fig.\ \ref{fig:fig5}. In this case, the flux periodicity is doubled due to the inclusion of paths encircling a single $AB$ or $BA$ triangle. In the presence of disorder, e.g.\ due to variations in the twist angle or charge density, the A-B resonances are broadened \cite{Note1}.

Our findings in the quasi-1D regime exhibit the same qualitative features as the experiment in Ref.\ \onlinecite{Xu2019} where resonances in the magnetoresistivity were observed at integer fractions of the main period on top of a plateau slightly above $h/8e^2 \approx 3.2$~k$\Omega$. This increase might be attributed to small nonzero $\phi$ or a different transport direction with respect to Fig.\ \ref{fig:fig3}(a). However, in Ref.\ \onlinecite{Xu2019} the main A-B period was interpreted in terms of interfering paths encircling a single $AB$ or $BA$ triangle. In our model, such paths are only allowed in the 2D regime and do not give rise to robust A-B resonances, as paths encircling an odd number of $AB$ or $BA$ triangles pick up a relative dynamical phase and their contributions are therefore smeared out at finite temperatures, as shown in Fig.\ \ref{fig:fig5}. This discrepancy can be resolved if instead the observed main period corresponds to paths encircling a moir\'e cell. This would be consistent with the observation of A-B resonances over a wide range of temperatures below the gap, and is equivalent to an increase of the twist angle reported in Ref.\ \onlinecite{Xu2019} by a factor $\sqrt{2}$. This ambiguity in the determination of the twist angle was already reported in earlier experimental works \cite{Cao2018a}. 

\emph{Conclusions} ---  We constructed a phenomenological scattering theory for the triangular network of valley Hall states that arises in minimally twisted bilayer graphene under interlayer bias, based solely on the symmetries of twisted bilayer graphene. In the absence of forward scattering, we showed that the network model depends only on the phase shift after intrachannel deflections, which tunes the system between one-dimensional chiral zigzag modes and pseudo-Landau levels. In this sense, our theory unifies these two phenomena, showing that both arise from the network. We then explored the effect of forward scattering between valley Hall states and discussed different coupling mechanisms between zigzag modes, which have important implications on the nature of magnetostransport oscillations.

We found that there are two transport regimes characterized by the effective dimensionality of the network, depending on the coupling regime. In the quasi-1D regime, only parallel zigzag channels are coupled which gives rise to Aharonov-Bohm oscillations in the magnetoconductance, whose periods correspond to threading one flux quantum through integer multiples of the moir\'e cell. Moreover, we demonstrated that these Aharonov-Bohm oscillations are independent of energy and hence they are robust against temperature. In contrast, in the 2D regime, zigzag channels propagating along different directions are coupled and the Aharonov-Bohm oscillations are accompanied by Shubnikov-de Haas oscillations. Remarkably, we found that the Aharonov-Bohm oscillations dominate at sufficiently high temperatures since the Shubnikov-de Haas oscillations are smeared out. Our findings agree well with experiments at twist angles $\theta \sim 0.1^\circ$ and low magnetic fields, and provide for the first time a transport theory for the topological network in minimally twisted bilayer graphene under interlayer bias.

\begin{acknowledgments}
\emph{Acknowledgments} --- F.D.\ and P.R.\ gratefully acknowledge funding by the Deutsche Forschungsgemeinschaft (DFG, German Research Foundation) within the framework of Germany's Excellence Strategy -- EXC-2123 QuantumFrontiers -- 390837967.
\end{acknowledgments}

\bibliography{references.bib}

%Control: production of eprint (-1) disabled
\begin{thebibliography}{40}%
\makeatletter
\providecommand \@ifxundefined [1]{%
 \@ifx{#1\undefined}
}%
\providecommand \@ifnum [1]{%
 \ifnum #1\expandafter \@firstoftwo
 \else \expandafter \@secondoftwo
 \fi
}%
\providecommand \@ifx [1]{%
 \ifx #1\expandafter \@firstoftwo
 \else \expandafter \@secondoftwo
 \fi
}%
\providecommand \natexlab [1]{#1}%
\providecommand \enquote  [1]{``#1''}%
\providecommand \bibnamefont  [1]{#1}%
\providecommand \bibfnamefont [1]{#1}%
\providecommand \citenamefont [1]{#1}%
\providecommand \href@noop [0]{\@secondoftwo}%
\providecommand \href [0]{\begingroup \@sanitize@url \@href}%
\providecommand \@href[1]{\@@startlink{#1}\@@href}%
\providecommand \@@href[1]{\endgroup#1\@@endlink}%
\providecommand \@sanitize@url [0]{\catcode `\\12\catcode `\$12\catcode
  `\&12\catcode `\#12\catcode `\^12\catcode `\_12\catcode `\%12\relax}%
\providecommand \@@startlink[1]{}%
\providecommand \@@endlink[0]{}%
\providecommand \url  [0]{\begingroup\@sanitize@url \@url }%
\providecommand \@url [1]{\endgroup\@href {#1}{\urlprefix }}%
\providecommand \urlprefix  [0]{URL }%
\providecommand \Eprint [0]{\href }%
\providecommand \doibase [0]{http://dx.doi.org/}%
\providecommand \selectlanguage [0]{\@gobble}%
\providecommand \bibinfo  [0]{\@secondoftwo}%
\providecommand \bibfield  [0]{\@secondoftwo}%
\providecommand \translation [1]{[#1]}%
\providecommand \BibitemOpen [0]{}%
\providecommand \bibitemStop [0]{}%
\providecommand \bibitemNoStop [0]{.\EOS\space}%
\providecommand \EOS [0]{\spacefactor3000\relax}%
\providecommand \BibitemShut  [1]{\csname bibitem#1\endcsname}%
\let\auto@bib@innerbib\@empty
%</preamble>
\bibitem [{\citenamefont {{Lopes dos Santos}}\ \emph
  {et~al.}(2007)\citenamefont {{Lopes dos Santos}}, \citenamefont {Peres},\
  and\ \citenamefont {{Castro Neto}}}]{LopesDosSantos2007}%
  \BibitemOpen
  \bibfield  {author} {\bibinfo {author} {\bibfnamefont {J.~M.~B.}\
  \bibnamefont {{Lopes dos Santos}}}, \bibinfo {author} {\bibfnamefont
  {N.~M.~R.}\ \bibnamefont {Peres}}, \ and\ \bibinfo {author} {\bibfnamefont
  {A.~H.}\ \bibnamefont {{Castro Neto}}},\ }\href {\doibase
  10.1103/PhysRevLett.99.256802} {\bibfield  {journal} {\bibinfo  {journal}
  {Phys. Rev. Lett.}\ }\textbf {\bibinfo {volume} {99}},\ \bibinfo {pages}
  {256802} (\bibinfo {year} {2007})}\BibitemShut {NoStop}%
\bibitem [{\citenamefont {{Su{\'{a}}rez Morell}}\ \emph
  {et~al.}(2010)\citenamefont {{Su{\'{a}}rez Morell}}, \citenamefont {Correa},
  \citenamefont {Vargas}, \citenamefont {Pacheco},\ and\ \citenamefont
  {Barticevic}}]{SuarezMorell2010}%
  \BibitemOpen
  \bibfield  {author} {\bibinfo {author} {\bibfnamefont {E.}~\bibnamefont
  {{Su{\'{a}}rez Morell}}}, \bibinfo {author} {\bibfnamefont {J.~D.}\
  \bibnamefont {Correa}}, \bibinfo {author} {\bibfnamefont {P.}~\bibnamefont
  {Vargas}}, \bibinfo {author} {\bibfnamefont {M.}~\bibnamefont {Pacheco}}, \
  and\ \bibinfo {author} {\bibfnamefont {Z.}~\bibnamefont {Barticevic}},\
  }\href {\doibase 10.1103/PhysRevB.82.121407} {\bibfield  {journal} {\bibinfo
  {journal} {Phys. Rev. B}\ }\textbf {\bibinfo {volume} {82}},\ \bibinfo
  {pages} {121407(R)} (\bibinfo {year} {2010})}\BibitemShut {NoStop}%
\bibitem [{\citenamefont {Bistritzer}\ and\ \citenamefont
  {MacDonald}(2011)}]{Bistritzer2010}%
  \BibitemOpen
  \bibfield  {author} {\bibinfo {author} {\bibfnamefont {R.}~\bibnamefont
  {Bistritzer}}\ and\ \bibinfo {author} {\bibfnamefont {A.~H.}\ \bibnamefont
  {MacDonald}},\ }\href {\doibase 10.1073/pnas.1108174108} {\bibfield
  {journal} {\bibinfo  {journal} {Proc. Natl. Acad. Sci.}\ }\textbf {\bibinfo
  {volume} {108}},\ \bibinfo {pages} {12233} (\bibinfo {year}
  {2011})}\BibitemShut {NoStop}%
\bibitem [{\citenamefont {Li}\ \emph {et~al.}(2010)\citenamefont {Li},
  \citenamefont {Luican}, \citenamefont {{Lopes dos Santos}}, \citenamefont
  {{Castro Neto}}, \citenamefont {Reina}, \citenamefont {Kong},\ and\
  \citenamefont {Andrei}}]{Li2010}%
  \BibitemOpen
  \bibfield  {author} {\bibinfo {author} {\bibfnamefont {G.}~\bibnamefont
  {Li}}, \bibinfo {author} {\bibfnamefont {A.}~\bibnamefont {Luican}}, \bibinfo
  {author} {\bibfnamefont {J.~M.~B.}\ \bibnamefont {{Lopes dos Santos}}},
  \bibinfo {author} {\bibfnamefont {A.~H.}\ \bibnamefont {{Castro Neto}}},
  \bibinfo {author} {\bibfnamefont {A.}~\bibnamefont {Reina}}, \bibinfo
  {author} {\bibfnamefont {J.}~\bibnamefont {Kong}}, \ and\ \bibinfo {author}
  {\bibfnamefont {E.~Y.}\ \bibnamefont {Andrei}},\ }\href {\doibase
  10.1038/nphys1463} {\bibfield  {journal} {\bibinfo  {journal} {Nat. Phys.}\
  }\textbf {\bibinfo {volume} {6}},\ \bibinfo {pages} {109} (\bibinfo {year}
  {2010})}\BibitemShut {NoStop}%
\bibitem [{\citenamefont {Kim}\ \emph {et~al.}(2017)\citenamefont {Kim},
  \citenamefont {DaSilva}, \citenamefont {Huang}, \citenamefont {Fallahazad},
  \citenamefont {Larentis}, \citenamefont {Taniguchi}, \citenamefont
  {Watanabe}, \citenamefont {LeRoy}, \citenamefont {MacDonald},\ and\
  \citenamefont {Tutuc}}]{Kim2017a}%
  \BibitemOpen
  \bibfield  {author} {\bibinfo {author} {\bibfnamefont {K.}~\bibnamefont
  {Kim}}, \bibinfo {author} {\bibfnamefont {A.}~\bibnamefont {DaSilva}},
  \bibinfo {author} {\bibfnamefont {S.}~\bibnamefont {Huang}}, \bibinfo
  {author} {\bibfnamefont {B.}~\bibnamefont {Fallahazad}}, \bibinfo {author}
  {\bibfnamefont {S.}~\bibnamefont {Larentis}}, \bibinfo {author}
  {\bibfnamefont {T.}~\bibnamefont {Taniguchi}}, \bibinfo {author}
  {\bibfnamefont {K.}~\bibnamefont {Watanabe}}, \bibinfo {author}
  {\bibfnamefont {B.~J.}\ \bibnamefont {LeRoy}}, \bibinfo {author}
  {\bibfnamefont {A.~H.}\ \bibnamefont {MacDonald}}, \ and\ \bibinfo {author}
  {\bibfnamefont {E.}~\bibnamefont {Tutuc}},\ }\href {\doibase
  10.1073/pnas.1620140114} {\bibfield  {journal} {\bibinfo  {journal} {Proc.
  Natl. Acad. Sci.}\ }\textbf {\bibinfo {volume} {114}},\ \bibinfo {pages}
  {3364} (\bibinfo {year} {2017})}\BibitemShut {NoStop}%
\bibitem [{\citenamefont {Cao}\ \emph {et~al.}(2018{\natexlab{a}})\citenamefont
  {Cao}, \citenamefont {Fatemi}, \citenamefont {Demir}, \citenamefont {Fang},
  \citenamefont {Tomarken}, \citenamefont {Luo}, \citenamefont
  {Sanchez-Yamagishi}, \citenamefont {Watanabe}, \citenamefont {Taniguchi},
  \citenamefont {Kaxiras}, \citenamefont {Ashoori},\ and\ \citenamefont
  {Jarillo-Herrero}}]{Cao2018a}%
  \BibitemOpen
  \bibfield  {author} {\bibinfo {author} {\bibfnamefont {Y.}~\bibnamefont
  {Cao}}, \bibinfo {author} {\bibfnamefont {V.}~\bibnamefont {Fatemi}},
  \bibinfo {author} {\bibfnamefont {A.}~\bibnamefont {Demir}}, \bibinfo
  {author} {\bibfnamefont {S.}~\bibnamefont {Fang}}, \bibinfo {author}
  {\bibfnamefont {S.~L.}\ \bibnamefont {Tomarken}}, \bibinfo {author}
  {\bibfnamefont {J.~Y.}\ \bibnamefont {Luo}}, \bibinfo {author} {\bibfnamefont
  {J.~D.}\ \bibnamefont {Sanchez-Yamagishi}}, \bibinfo {author} {\bibfnamefont
  {K.}~\bibnamefont {Watanabe}}, \bibinfo {author} {\bibfnamefont
  {T.}~\bibnamefont {Taniguchi}}, \bibinfo {author} {\bibfnamefont
  {E.}~\bibnamefont {Kaxiras}}, \bibinfo {author} {\bibfnamefont {R.~C.}\
  \bibnamefont {Ashoori}}, \ and\ \bibinfo {author} {\bibfnamefont
  {P.}~\bibnamefont {Jarillo-Herrero}},\ }\href {\doibase 10.1038/nature26154}
  {\bibfield  {journal} {\bibinfo  {journal} {Nature}\ }\textbf {\bibinfo
  {volume} {556}},\ \bibinfo {pages} {80} (\bibinfo {year}
  {2018}{\natexlab{a}})}\BibitemShut {NoStop}%
\bibitem [{\citenamefont {Cao}\ \emph {et~al.}(2018{\natexlab{b}})\citenamefont
  {Cao}, \citenamefont {Fatemi}, \citenamefont {Fang}, \citenamefont
  {Watanabe}, \citenamefont {Taniguchi}, \citenamefont {Kaxiras},\ and\
  \citenamefont {Jarillo-Herrero}}]{Cao2018}%
  \BibitemOpen
  \bibfield  {author} {\bibinfo {author} {\bibfnamefont {Y.}~\bibnamefont
  {Cao}}, \bibinfo {author} {\bibfnamefont {V.}~\bibnamefont {Fatemi}},
  \bibinfo {author} {\bibfnamefont {S.}~\bibnamefont {Fang}}, \bibinfo {author}
  {\bibfnamefont {K.}~\bibnamefont {Watanabe}}, \bibinfo {author}
  {\bibfnamefont {T.}~\bibnamefont {Taniguchi}}, \bibinfo {author}
  {\bibfnamefont {E.}~\bibnamefont {Kaxiras}}, \ and\ \bibinfo {author}
  {\bibfnamefont {P.}~\bibnamefont {Jarillo-Herrero}},\ }\href {\doibase
  10.1038/nature26160} {\bibfield  {journal} {\bibinfo  {journal} {Nature}\
  }\textbf {\bibinfo {volume} {556}},\ \bibinfo {pages} {43} (\bibinfo {year}
  {2018}{\natexlab{b}})}\BibitemShut {NoStop}%
\bibitem [{\citenamefont {Yankowitz}\ \emph {et~al.}(2019)\citenamefont
  {Yankowitz}, \citenamefont {Chen}, \citenamefont {Polshyn}, \citenamefont
  {Zhang}, \citenamefont {Watanabe}, \citenamefont {Taniguchi}, \citenamefont
  {Graf}, \citenamefont {Young},\ and\ \citenamefont {Dean}}]{Yankowitz2019}%
  \BibitemOpen
  \bibfield  {author} {\bibinfo {author} {\bibfnamefont {M.}~\bibnamefont
  {Yankowitz}}, \bibinfo {author} {\bibfnamefont {S.}~\bibnamefont {Chen}},
  \bibinfo {author} {\bibfnamefont {H.}~\bibnamefont {Polshyn}}, \bibinfo
  {author} {\bibfnamefont {Y.}~\bibnamefont {Zhang}}, \bibinfo {author}
  {\bibfnamefont {K.}~\bibnamefont {Watanabe}}, \bibinfo {author}
  {\bibfnamefont {T.}~\bibnamefont {Taniguchi}}, \bibinfo {author}
  {\bibfnamefont {D.}~\bibnamefont {Graf}}, \bibinfo {author} {\bibfnamefont
  {A.~F.}\ \bibnamefont {Young}}, \ and\ \bibinfo {author} {\bibfnamefont
  {C.~R.}\ \bibnamefont {Dean}},\ }\href {\doibase 10.1126/science.aav1910}
  {\bibfield  {journal} {\bibinfo  {journal} {Science}\ }\textbf {\bibinfo
  {volume} {363}},\ \bibinfo {pages} {1059} (\bibinfo {year}
  {2019})}\BibitemShut {NoStop}%
\bibitem [{\citenamefont {Sharpe}\ \emph {et~al.}(2019)\citenamefont {Sharpe},
  \citenamefont {Fox}, \citenamefont {Barnard}, \citenamefont {Finney},
  \citenamefont {Watanabe}, \citenamefont {Taniguchi}, \citenamefont
  {Kastner},\ and\ \citenamefont {Goldhaber-Gordon}}]{Sharpe2019}%
  \BibitemOpen
  \bibfield  {author} {\bibinfo {author} {\bibfnamefont {A.~L.}\ \bibnamefont
  {Sharpe}}, \bibinfo {author} {\bibfnamefont {E.~J.}\ \bibnamefont {Fox}},
  \bibinfo {author} {\bibfnamefont {A.~W.}\ \bibnamefont {Barnard}}, \bibinfo
  {author} {\bibfnamefont {J.}~\bibnamefont {Finney}}, \bibinfo {author}
  {\bibfnamefont {K.}~\bibnamefont {Watanabe}}, \bibinfo {author}
  {\bibfnamefont {T.}~\bibnamefont {Taniguchi}}, \bibinfo {author}
  {\bibfnamefont {M.~A.}\ \bibnamefont {Kastner}}, \ and\ \bibinfo {author}
  {\bibfnamefont {D.}~\bibnamefont {Goldhaber-Gordon}},\ }\href {\doibase
  10.1126/science.aaw3780} {\bibfield  {journal} {\bibinfo  {journal}
  {Science}\ }\textbf {\bibinfo {volume} {365}},\ \bibinfo {pages} {605}
  (\bibinfo {year} {2019})}\BibitemShut {NoStop}%
\bibitem [{\citenamefont {Kerelsky}\ \emph {et~al.}(2019)\citenamefont
  {Kerelsky}, \citenamefont {McGilly}, \citenamefont {Kennes}, \citenamefont
  {Xian}, \citenamefont {Yankowitz}, \citenamefont {Chen}, \citenamefont
  {Watanabe}, \citenamefont {Taniguchi}, \citenamefont {Hone}, \citenamefont
  {Dean}, \citenamefont {Rubio},\ and\ \citenamefont
  {Pasupathy}}]{Kerelsky2019}%
  \BibitemOpen
  \bibfield  {author} {\bibinfo {author} {\bibfnamefont {A.}~\bibnamefont
  {Kerelsky}}, \bibinfo {author} {\bibfnamefont {L.~J.}\ \bibnamefont
  {McGilly}}, \bibinfo {author} {\bibfnamefont {D.~M.}\ \bibnamefont {Kennes}},
  \bibinfo {author} {\bibfnamefont {L.}~\bibnamefont {Xian}}, \bibinfo {author}
  {\bibfnamefont {M.}~\bibnamefont {Yankowitz}}, \bibinfo {author}
  {\bibfnamefont {S.}~\bibnamefont {Chen}}, \bibinfo {author} {\bibfnamefont
  {K.}~\bibnamefont {Watanabe}}, \bibinfo {author} {\bibfnamefont
  {T.}~\bibnamefont {Taniguchi}}, \bibinfo {author} {\bibfnamefont
  {J.}~\bibnamefont {Hone}}, \bibinfo {author} {\bibfnamefont {C.}~\bibnamefont
  {Dean}}, \bibinfo {author} {\bibfnamefont {A.}~\bibnamefont {Rubio}}, \ and\
  \bibinfo {author} {\bibfnamefont {A.~N.}\ \bibnamefont {Pasupathy}},\ }\href
  {\doibase 10.1038/s41586-019-1431-9} {\bibfield  {journal} {\bibinfo
  {journal} {Nature}\ }\textbf {\bibinfo {volume} {572}},\ \bibinfo {pages}
  {95} (\bibinfo {year} {2019})}\BibitemShut {NoStop}%
\bibitem [{\citenamefont {Choi}\ \emph {et~al.}(2019)\citenamefont {Choi},
  \citenamefont {Kemmer}, \citenamefont {Peng}, \citenamefont {Thomson},
  \citenamefont {Arora}, \citenamefont {Polski}, \citenamefont {Zhang},
  \citenamefont {Ren}, \citenamefont {Alicea}, \citenamefont {Refael},
  \citenamefont {von Oppen}, \citenamefont {Watanabe}, \citenamefont
  {Taniguchi},\ and\ \citenamefont {Nadj-Perge}}]{Choi2019}%
  \BibitemOpen
  \bibfield  {author} {\bibinfo {author} {\bibfnamefont {Y.}~\bibnamefont
  {Choi}}, \bibinfo {author} {\bibfnamefont {J.}~\bibnamefont {Kemmer}},
  \bibinfo {author} {\bibfnamefont {Y.}~\bibnamefont {Peng}}, \bibinfo {author}
  {\bibfnamefont {A.}~\bibnamefont {Thomson}}, \bibinfo {author} {\bibfnamefont
  {H.}~\bibnamefont {Arora}}, \bibinfo {author} {\bibfnamefont
  {R.}~\bibnamefont {Polski}}, \bibinfo {author} {\bibfnamefont
  {Y.}~\bibnamefont {Zhang}}, \bibinfo {author} {\bibfnamefont
  {H.}~\bibnamefont {Ren}}, \bibinfo {author} {\bibfnamefont {J.}~\bibnamefont
  {Alicea}}, \bibinfo {author} {\bibfnamefont {G.}~\bibnamefont {Refael}},
  \bibinfo {author} {\bibfnamefont {F.}~\bibnamefont {von Oppen}}, \bibinfo
  {author} {\bibfnamefont {K.}~\bibnamefont {Watanabe}}, \bibinfo {author}
  {\bibfnamefont {T.}~\bibnamefont {Taniguchi}}, \ and\ \bibinfo {author}
  {\bibfnamefont {S.}~\bibnamefont {Nadj-Perge}},\ }\href {\doibase
  10.1038/s41567-019-0606-5} {\bibfield  {journal} {\bibinfo  {journal} {Nat.
  Phys.}\ }\textbf {\bibinfo {volume} {15}},\ \bibinfo {pages} {1174} (\bibinfo
  {year} {2019})}\BibitemShut {NoStop}%
\bibitem [{\citenamefont {Cao}\ \emph {et~al.}(2020)\citenamefont {Cao},
  \citenamefont {Chowdhury}, \citenamefont {Rodan-Legrain}, \citenamefont
  {Rubies-Bigord{\`{a}}}, \citenamefont {Watanabe}, \citenamefont {Taniguchi},
  \citenamefont {Senthil},\ and\ \citenamefont {Jarillo-Herrero}}]{Cao2019}%
  \BibitemOpen
  \bibfield  {author} {\bibinfo {author} {\bibfnamefont {Y.}~\bibnamefont
  {Cao}}, \bibinfo {author} {\bibfnamefont {D.}~\bibnamefont {Chowdhury}},
  \bibinfo {author} {\bibfnamefont {D.}~\bibnamefont {Rodan-Legrain}}, \bibinfo
  {author} {\bibfnamefont {O.}~\bibnamefont {Rubies-Bigord{\`{a}}}}, \bibinfo
  {author} {\bibfnamefont {K.}~\bibnamefont {Watanabe}}, \bibinfo {author}
  {\bibfnamefont {T.}~\bibnamefont {Taniguchi}}, \bibinfo {author}
  {\bibfnamefont {T.}~\bibnamefont {Senthil}}, \ and\ \bibinfo {author}
  {\bibfnamefont {P.}~\bibnamefont {Jarillo-Herrero}},\ }\href {\doibase
  10.1103/PhysRevLett.124.076801} {\bibfield  {journal} {\bibinfo  {journal}
  {Phys. Rev. Lett.}\ }\textbf {\bibinfo {volume} {124}},\ \bibinfo {pages}
  {76801} (\bibinfo {year} {2020})}\BibitemShut {NoStop}%
\bibitem [{\citenamefont {Nam}\ and\ \citenamefont {Koshino}(2017)}]{Nam2017}%
  \BibitemOpen
  \bibfield  {author} {\bibinfo {author} {\bibfnamefont {N.~N.~T.}\
  \bibnamefont {Nam}}\ and\ \bibinfo {author} {\bibfnamefont {M.}~\bibnamefont
  {Koshino}},\ }\href {\doibase 10.1103/PhysRevB.96.075311} {\bibfield
  {journal} {\bibinfo  {journal} {Phys. Rev. B}\ }\textbf {\bibinfo {volume}
  {96}},\ \bibinfo {pages} {075311} (\bibinfo {year} {2017})}\BibitemShut
  {NoStop}%
\bibitem [{\citenamefont {Yoo}\ \emph {et~al.}(2019)\citenamefont {Yoo},
  \citenamefont {Engelke}, \citenamefont {Carr}, \citenamefont {Fang},
  \citenamefont {Zhang}, \citenamefont {Cazeaux}, \citenamefont {Sung},
  \citenamefont {Hovden}, \citenamefont {Tsen}, \citenamefont {Taniguchi},
  \citenamefont {Watanabe}, \citenamefont {Yi}, \citenamefont {Kim},
  \citenamefont {Luskin}, \citenamefont {Tadmor}, \citenamefont {Kaxiras},\
  and\ \citenamefont {Kim}}]{Yoo2019}%
  \BibitemOpen
  \bibfield  {author} {\bibinfo {author} {\bibfnamefont {H.}~\bibnamefont
  {Yoo}}, \bibinfo {author} {\bibfnamefont {R.}~\bibnamefont {Engelke}},
  \bibinfo {author} {\bibfnamefont {S.}~\bibnamefont {Carr}}, \bibinfo {author}
  {\bibfnamefont {S.}~\bibnamefont {Fang}}, \bibinfo {author} {\bibfnamefont
  {K.}~\bibnamefont {Zhang}}, \bibinfo {author} {\bibfnamefont
  {P.}~\bibnamefont {Cazeaux}}, \bibinfo {author} {\bibfnamefont {S.~H.}\
  \bibnamefont {Sung}}, \bibinfo {author} {\bibfnamefont {R.}~\bibnamefont
  {Hovden}}, \bibinfo {author} {\bibfnamefont {A.~W.}\ \bibnamefont {Tsen}},
  \bibinfo {author} {\bibfnamefont {T.}~\bibnamefont {Taniguchi}}, \bibinfo
  {author} {\bibfnamefont {K.}~\bibnamefont {Watanabe}}, \bibinfo {author}
  {\bibfnamefont {G.-C.}\ \bibnamefont {Yi}}, \bibinfo {author} {\bibfnamefont
  {M.}~\bibnamefont {Kim}}, \bibinfo {author} {\bibfnamefont {M.}~\bibnamefont
  {Luskin}}, \bibinfo {author} {\bibfnamefont {E.~B.}\ \bibnamefont {Tadmor}},
  \bibinfo {author} {\bibfnamefont {E.}~\bibnamefont {Kaxiras}}, \ and\
  \bibinfo {author} {\bibfnamefont {P.}~\bibnamefont {Kim}},\ }\href {\doibase
  10.1038/s41563-019-0346-z} {\bibfield  {journal} {\bibinfo  {journal} {Nat.
  Mater.}\ }\textbf {\bibinfo {volume} {18}},\ \bibinfo {pages} {448} (\bibinfo
  {year} {2019})}\BibitemShut {NoStop}%
\bibitem [{\citenamefont {Walet}\ and\ \citenamefont
  {Guinea}(2019)}]{Walet2019}%
  \BibitemOpen
  \bibfield  {author} {\bibinfo {author} {\bibfnamefont {N.~R.}\ \bibnamefont
  {Walet}}\ and\ \bibinfo {author} {\bibfnamefont {F.}~\bibnamefont {Guinea}},\
  }\href {\doibase 10.1088/2053-1583/ab57f8} {\bibfield  {journal} {\bibinfo
  {journal} {2D Mater.}\ }\textbf {\bibinfo {volume} {7}},\ \bibinfo {pages}
  {015023} (\bibinfo {year} {2019})}\BibitemShut {NoStop}%
\bibitem [{\citenamefont {Martin}\ \emph {et~al.}(2008)\citenamefont {Martin},
  \citenamefont {Blanter},\ and\ \citenamefont {Morpurgo}}]{Martin2008}%
  \BibitemOpen
  \bibfield  {author} {\bibinfo {author} {\bibfnamefont {I.}~\bibnamefont
  {Martin}}, \bibinfo {author} {\bibfnamefont {Y.~M.}\ \bibnamefont {Blanter}},
  \ and\ \bibinfo {author} {\bibfnamefont {A.~F.}\ \bibnamefont {Morpurgo}},\
  }\href {\doibase 10.1103/PhysRevLett.100.036804} {\bibfield  {journal}
  {\bibinfo  {journal} {Phys. Rev. Lett.}\ }\textbf {\bibinfo {volume} {100}},\
  \bibinfo {pages} {036804} (\bibinfo {year} {2008})}\BibitemShut {NoStop}%
\bibitem [{\citenamefont {Zhang}\ \emph {et~al.}(2013)\citenamefont {Zhang},
  \citenamefont {MacDonald},\ and\ \citenamefont {Mele}}]{Zhang2013}%
  \BibitemOpen
  \bibfield  {author} {\bibinfo {author} {\bibfnamefont {F.}~\bibnamefont
  {Zhang}}, \bibinfo {author} {\bibfnamefont {A.~H.}\ \bibnamefont
  {MacDonald}}, \ and\ \bibinfo {author} {\bibfnamefont {E.~J.}\ \bibnamefont
  {Mele}},\ }\href {\doibase 10.1073/pnas.1308853110} {\bibfield  {journal}
  {\bibinfo  {journal} {Proc. Natl. Acad. Sci.}\ }\textbf {\bibinfo {volume}
  {110}},\ \bibinfo {pages} {10546} (\bibinfo {year} {2013})}\BibitemShut
  {NoStop}%
\bibitem [{\citenamefont {Vaezi}\ \emph {et~al.}(2013)\citenamefont {Vaezi},
  \citenamefont {Liang}, \citenamefont {Ngai}, \citenamefont {Yang},\ and\
  \citenamefont {Kim}}]{Vaezi2013}%
  \BibitemOpen
  \bibfield  {author} {\bibinfo {author} {\bibfnamefont {A.}~\bibnamefont
  {Vaezi}}, \bibinfo {author} {\bibfnamefont {Y.}~\bibnamefont {Liang}},
  \bibinfo {author} {\bibfnamefont {D.~H.}\ \bibnamefont {Ngai}}, \bibinfo
  {author} {\bibfnamefont {L.}~\bibnamefont {Yang}}, \ and\ \bibinfo {author}
  {\bibfnamefont {E.-A.}\ \bibnamefont {Kim}},\ }\href {\doibase
  10.1103/PhysRevX.3.021018} {\bibfield  {journal} {\bibinfo  {journal} {Phys.
  Rev. X}\ }\textbf {\bibinfo {volume} {3}},\ \bibinfo {pages} {021018}
  (\bibinfo {year} {2013})}\BibitemShut {NoStop}%
\bibitem [{\citenamefont {Ju}\ \emph {et~al.}(2015)\citenamefont {Ju},
  \citenamefont {Shi}, \citenamefont {Nair}, \citenamefont {Lv}, \citenamefont
  {Jin}, \citenamefont {Velasco}, \citenamefont {Ojeda-Aristizabal},
  \citenamefont {Bechtel}, \citenamefont {Martin}, \citenamefont {Zettl},
  \citenamefont {Analytis},\ and\ \citenamefont {Wang}}]{Ju2015}%
  \BibitemOpen
  \bibfield  {author} {\bibinfo {author} {\bibfnamefont {L.}~\bibnamefont
  {Ju}}, \bibinfo {author} {\bibfnamefont {Z.}~\bibnamefont {Shi}}, \bibinfo
  {author} {\bibfnamefont {N.}~\bibnamefont {Nair}}, \bibinfo {author}
  {\bibfnamefont {Y.}~\bibnamefont {Lv}}, \bibinfo {author} {\bibfnamefont
  {C.}~\bibnamefont {Jin}}, \bibinfo {author} {\bibfnamefont {J.}~\bibnamefont
  {Velasco}}, \bibinfo {author} {\bibfnamefont {C.}~\bibnamefont
  {Ojeda-Aristizabal}}, \bibinfo {author} {\bibfnamefont {H.~A.}\ \bibnamefont
  {Bechtel}}, \bibinfo {author} {\bibfnamefont {M.~C.}\ \bibnamefont {Martin}},
  \bibinfo {author} {\bibfnamefont {A.}~\bibnamefont {Zettl}}, \bibinfo
  {author} {\bibfnamefont {J.}~\bibnamefont {Analytis}}, \ and\ \bibinfo
  {author} {\bibfnamefont {F.}~\bibnamefont {Wang}},\ }\href {\doibase
  10.1038/nature14364} {\bibfield  {journal} {\bibinfo  {journal} {Nature}\
  }\textbf {\bibinfo {volume} {520}},\ \bibinfo {pages} {650} (\bibinfo {year}
  {2015})}\BibitemShut {NoStop}%
\bibitem [{\citenamefont {Yin}\ \emph {et~al.}(2016)\citenamefont {Yin},
  \citenamefont {Jiang}, \citenamefont {Qiao},\ and\ \citenamefont
  {He}}]{Yin2016}%
  \BibitemOpen
  \bibfield  {author} {\bibinfo {author} {\bibfnamefont {L.-J.}\ \bibnamefont
  {Yin}}, \bibinfo {author} {\bibfnamefont {H.}~\bibnamefont {Jiang}}, \bibinfo
  {author} {\bibfnamefont {J.-B.}\ \bibnamefont {Qiao}}, \ and\ \bibinfo
  {author} {\bibfnamefont {L.}~\bibnamefont {He}},\ }\href {\doibase
  10.1038/ncomms11760} {\bibfield  {journal} {\bibinfo  {journal} {Nat.
  Commun.}\ }\textbf {\bibinfo {volume} {7}},\ \bibinfo {pages} {11760}
  (\bibinfo {year} {2016})}\BibitemShut {NoStop}%
\bibitem [{\citenamefont {San-Jose}\ and\ \citenamefont
  {Prada}(2013)}]{San-jose2013}%
  \BibitemOpen
  \bibfield  {author} {\bibinfo {author} {\bibfnamefont {P.}~\bibnamefont
  {San-Jose}}\ and\ \bibinfo {author} {\bibfnamefont {E.}~\bibnamefont
  {Prada}},\ }\href {\doibase 10.1103/PhysRevB.88.121408} {\bibfield  {journal}
  {\bibinfo  {journal} {Phys. Rev. B}\ }\textbf {\bibinfo {volume} {88}},\
  \bibinfo {pages} {121408(R)} (\bibinfo {year} {2013})}\BibitemShut {NoStop}%
\bibitem [{\citenamefont {Efimkin}\ and\ \citenamefont
  {MacDonald}(2018)}]{Efimkin2018}%
  \BibitemOpen
  \bibfield  {author} {\bibinfo {author} {\bibfnamefont {D.~K.}\ \bibnamefont
  {Efimkin}}\ and\ \bibinfo {author} {\bibfnamefont {A.~H.}\ \bibnamefont
  {MacDonald}},\ }\href {\doibase 10.1103/PhysRevB.98.035404} {\bibfield
  {journal} {\bibinfo  {journal} {Phys. Rev. B}\ }\textbf {\bibinfo {volume}
  {98}},\ \bibinfo {pages} {035404} (\bibinfo {year} {2018})}\BibitemShut
  {NoStop}%
\bibitem [{\citenamefont {Ramires}\ and\ \citenamefont
  {Lado}(2018)}]{Ramires2018}%
  \BibitemOpen
  \bibfield  {author} {\bibinfo {author} {\bibfnamefont {A.}~\bibnamefont
  {Ramires}}\ and\ \bibinfo {author} {\bibfnamefont {J.~L.}\ \bibnamefont
  {Lado}},\ }\href {\doibase 10.1103/PhysRevLett.121.146801} {\bibfield
  {journal} {\bibinfo  {journal} {Phys. Rev. Lett.}\ }\textbf {\bibinfo
  {volume} {121}},\ \bibinfo {pages} {146801} (\bibinfo {year}
  {2018})}\BibitemShut {NoStop}%
\bibitem [{\citenamefont {Huang}\ \emph {et~al.}(2018)\citenamefont {Huang},
  \citenamefont {Kim}, \citenamefont {Efimkin}, \citenamefont {Lovorn},
  \citenamefont {Taniguchi}, \citenamefont {Watanabe}, \citenamefont
  {MacDonald}, \citenamefont {Tutuc},\ and\ \citenamefont {LeRoy}}]{Huang2018}%
  \BibitemOpen
  \bibfield  {author} {\bibinfo {author} {\bibfnamefont {S.}~\bibnamefont
  {Huang}}, \bibinfo {author} {\bibfnamefont {K.}~\bibnamefont {Kim}}, \bibinfo
  {author} {\bibfnamefont {D.~K.}\ \bibnamefont {Efimkin}}, \bibinfo {author}
  {\bibfnamefont {T.}~\bibnamefont {Lovorn}}, \bibinfo {author} {\bibfnamefont
  {T.}~\bibnamefont {Taniguchi}}, \bibinfo {author} {\bibfnamefont
  {K.}~\bibnamefont {Watanabe}}, \bibinfo {author} {\bibfnamefont {A.~H.}\
  \bibnamefont {MacDonald}}, \bibinfo {author} {\bibfnamefont {E.}~\bibnamefont
  {Tutuc}}, \ and\ \bibinfo {author} {\bibfnamefont {B.~J.}\ \bibnamefont
  {LeRoy}},\ }\href {\doibase 10.1103/PhysRevLett.121.037702} {\bibfield
  {journal} {\bibinfo  {journal} {Phys. Rev. Lett.}\ }\textbf {\bibinfo
  {volume} {121}},\ \bibinfo {pages} {037702} (\bibinfo {year}
  {2018})}\BibitemShut {NoStop}%
\bibitem [{\citenamefont {Sunku}\ \emph {et~al.}(2018)\citenamefont {Sunku},
  \citenamefont {Ni}, \citenamefont {Jiang}, \citenamefont {Yoo}, \citenamefont
  {Sternbach}, \citenamefont {McLeod}, \citenamefont {Stauber}, \citenamefont
  {Xiong}, \citenamefont {Taniguchi}, \citenamefont {Watanabe}, \citenamefont
  {Kim}, \citenamefont {Fogler},\ and\ \citenamefont {Basov}}]{Sunku2018}%
  \BibitemOpen
  \bibfield  {author} {\bibinfo {author} {\bibfnamefont {S.~S.}\ \bibnamefont
  {Sunku}}, \bibinfo {author} {\bibfnamefont {G.~X.}\ \bibnamefont {Ni}},
  \bibinfo {author} {\bibfnamefont {B.~Y.}\ \bibnamefont {Jiang}}, \bibinfo
  {author} {\bibfnamefont {H.}~\bibnamefont {Yoo}}, \bibinfo {author}
  {\bibfnamefont {A.}~\bibnamefont {Sternbach}}, \bibinfo {author}
  {\bibfnamefont {A.~S.}\ \bibnamefont {McLeod}}, \bibinfo {author}
  {\bibfnamefont {T.}~\bibnamefont {Stauber}}, \bibinfo {author} {\bibfnamefont
  {L.}~\bibnamefont {Xiong}}, \bibinfo {author} {\bibfnamefont
  {T.}~\bibnamefont {Taniguchi}}, \bibinfo {author} {\bibfnamefont
  {K.}~\bibnamefont {Watanabe}}, \bibinfo {author} {\bibfnamefont
  {P.}~\bibnamefont {Kim}}, \bibinfo {author} {\bibfnamefont {M.~M.}\
  \bibnamefont {Fogler}}, \ and\ \bibinfo {author} {\bibfnamefont {D.~N.}\
  \bibnamefont {Basov}},\ }\href {\doibase 10.1126/science.aau5144} {\bibfield
  {journal} {\bibinfo  {journal} {Science}\ }\textbf {\bibinfo {volume}
  {362}},\ \bibinfo {pages} {1153} (\bibinfo {year} {2018})}\BibitemShut
  {NoStop}%
\bibitem [{\citenamefont {Fleischmann}\ \emph {et~al.}(2020)\citenamefont
  {Fleischmann}, \citenamefont {Gupta}, \citenamefont {Wullschl{\"{a}}ger},
  \citenamefont {Theil}, \citenamefont {Weckbecker}, \citenamefont {Meded},
  \citenamefont {Sharma}, \citenamefont {Meyer},\ and\ \citenamefont
  {Shallcross}}]{Fleischmann2020}%
  \BibitemOpen
  \bibfield  {author} {\bibinfo {author} {\bibfnamefont {M.}~\bibnamefont
  {Fleischmann}}, \bibinfo {author} {\bibfnamefont {R.}~\bibnamefont {Gupta}},
  \bibinfo {author} {\bibfnamefont {F.}~\bibnamefont {Wullschl{\"{a}}ger}},
  \bibinfo {author} {\bibfnamefont {S.}~\bibnamefont {Theil}}, \bibinfo
  {author} {\bibfnamefont {D.}~\bibnamefont {Weckbecker}}, \bibinfo {author}
  {\bibfnamefont {V.}~\bibnamefont {Meded}}, \bibinfo {author} {\bibfnamefont
  {S.}~\bibnamefont {Sharma}}, \bibinfo {author} {\bibfnamefont
  {B.}~\bibnamefont {Meyer}}, \ and\ \bibinfo {author} {\bibfnamefont
  {S.}~\bibnamefont {Shallcross}},\ }\href {\doibase
  10.1021/acs.nanolett.9b04027} {\bibfield  {journal} {\bibinfo  {journal}
  {Nano Lett.}\ }\textbf {\bibinfo {volume} {20}},\ \bibinfo {pages} {971}
  (\bibinfo {year} {2020})}\BibitemShut {NoStop}%
\bibitem [{\citenamefont {Tsim}\ \emph {et~al.}(2020)\citenamefont {Tsim},
  \citenamefont {Nam},\ and\ \citenamefont {Koshino}}]{Tsim2020}%
  \BibitemOpen
  \bibfield  {author} {\bibinfo {author} {\bibfnamefont {B.}~\bibnamefont
  {Tsim}}, \bibinfo {author} {\bibfnamefont {N.~N.~T.}\ \bibnamefont {Nam}}, \
  and\ \bibinfo {author} {\bibfnamefont {M.}~\bibnamefont {Koshino}},\ }\href
  {\doibase 10.1103/PhysRevB.101.125409} {\bibfield  {journal} {\bibinfo
  {journal} {Phys. Rev. B}\ }\textbf {\bibinfo {volume} {101}},\ \bibinfo
  {pages} {125409} (\bibinfo {year} {2020})}\BibitemShut {NoStop}%
\bibitem [{\citenamefont {Rickhaus}\ \emph {et~al.}(2018)\citenamefont
  {Rickhaus}, \citenamefont {Wallbank}, \citenamefont {Slizovskiy},
  \citenamefont {Pisoni}, \citenamefont {Overweg}, \citenamefont {Lee},
  \citenamefont {Eich}, \citenamefont {Liu}, \citenamefont {Watanabe},
  \citenamefont {Taniguchi}, \citenamefont {Ihn},\ and\ \citenamefont
  {Ensslin}}]{Rickhaus2018}%
  \BibitemOpen
  \bibfield  {author} {\bibinfo {author} {\bibfnamefont {P.}~\bibnamefont
  {Rickhaus}}, \bibinfo {author} {\bibfnamefont {J.}~\bibnamefont {Wallbank}},
  \bibinfo {author} {\bibfnamefont {S.}~\bibnamefont {Slizovskiy}}, \bibinfo
  {author} {\bibfnamefont {R.}~\bibnamefont {Pisoni}}, \bibinfo {author}
  {\bibfnamefont {H.}~\bibnamefont {Overweg}}, \bibinfo {author} {\bibfnamefont
  {Y.}~\bibnamefont {Lee}}, \bibinfo {author} {\bibfnamefont {M.}~\bibnamefont
  {Eich}}, \bibinfo {author} {\bibfnamefont {M.-H.}\ \bibnamefont {Liu}},
  \bibinfo {author} {\bibfnamefont {K.}~\bibnamefont {Watanabe}}, \bibinfo
  {author} {\bibfnamefont {T.}~\bibnamefont {Taniguchi}}, \bibinfo {author}
  {\bibfnamefont {T.}~\bibnamefont {Ihn}}, \ and\ \bibinfo {author}
  {\bibfnamefont {K.}~\bibnamefont {Ensslin}},\ }\href {\doibase
  10.1021/acs.nanolett.8b02387} {\bibfield  {journal} {\bibinfo  {journal}
  {Nano Lett.}\ }\textbf {\bibinfo {volume} {18}},\ \bibinfo {pages} {6725}
  (\bibinfo {year} {2018})}\BibitemShut {NoStop}%
\bibitem [{\citenamefont {Xu}\ \emph {et~al.}(2019)\citenamefont {Xu},
  \citenamefont {Berdyugin}, \citenamefont {Kumaravadivel}, \citenamefont
  {Guinea}, \citenamefont {{Krishna Kumar}}, \citenamefont {Bandurin},
  \citenamefont {Morozov}, \citenamefont {Kuang}, \citenamefont {Tsim},
  \citenamefont {Liu}, \citenamefont {Edgar}, \citenamefont {Grigorieva},
  \citenamefont {Fal'ko}, \citenamefont {Kim},\ and\ \citenamefont
  {Geim}}]{Xu2019}%
  \BibitemOpen
  \bibfield  {author} {\bibinfo {author} {\bibfnamefont {S.~G.}\ \bibnamefont
  {Xu}}, \bibinfo {author} {\bibfnamefont {A.~I.}\ \bibnamefont {Berdyugin}},
  \bibinfo {author} {\bibfnamefont {P.}~\bibnamefont {Kumaravadivel}}, \bibinfo
  {author} {\bibfnamefont {F.}~\bibnamefont {Guinea}}, \bibinfo {author}
  {\bibfnamefont {R.}~\bibnamefont {{Krishna Kumar}}}, \bibinfo {author}
  {\bibfnamefont {D.~A.}\ \bibnamefont {Bandurin}}, \bibinfo {author}
  {\bibfnamefont {S.~V.}\ \bibnamefont {Morozov}}, \bibinfo {author}
  {\bibfnamefont {W.}~\bibnamefont {Kuang}}, \bibinfo {author} {\bibfnamefont
  {B.}~\bibnamefont {Tsim}}, \bibinfo {author} {\bibfnamefont {S.}~\bibnamefont
  {Liu}}, \bibinfo {author} {\bibfnamefont {J.~H.}\ \bibnamefont {Edgar}},
  \bibinfo {author} {\bibfnamefont {I.~V.}\ \bibnamefont {Grigorieva}},
  \bibinfo {author} {\bibfnamefont {V.~I.}\ \bibnamefont {Fal'ko}}, \bibinfo
  {author} {\bibfnamefont {M.}~\bibnamefont {Kim}}, \ and\ \bibinfo {author}
  {\bibfnamefont {A.~K.}\ \bibnamefont {Geim}},\ }\href {\doibase
  10.1038/s41467-019-11971-7} {\bibfield  {journal} {\bibinfo  {journal} {Nat.
  Commun.}\ }\textbf {\bibinfo {volume} {10}},\ \bibinfo {pages} {4008}
  (\bibinfo {year} {2019})}\BibitemShut {NoStop}%
\bibitem [{\citenamefont {Ji}\ \emph {et~al.}(2003)\citenamefont {Ji},
  \citenamefont {Chung}, \citenamefont {Sprinzak}, \citenamefont {Heiblum},
  \citenamefont {Mahalu},\ and\ \citenamefont {Shtrikman}}]{Ji2003}%
  \BibitemOpen
  \bibfield  {author} {\bibinfo {author} {\bibfnamefont {Y.}~\bibnamefont
  {Ji}}, \bibinfo {author} {\bibfnamefont {Y.}~\bibnamefont {Chung}}, \bibinfo
  {author} {\bibfnamefont {D.}~\bibnamefont {Sprinzak}}, \bibinfo {author}
  {\bibfnamefont {M.}~\bibnamefont {Heiblum}}, \bibinfo {author} {\bibfnamefont
  {D.}~\bibnamefont {Mahalu}}, \ and\ \bibinfo {author} {\bibfnamefont
  {H.}~\bibnamefont {Shtrikman}},\ }\href {\doibase 10.1038/nature01503}
  {\bibfield  {journal} {\bibinfo  {journal} {Nature}\ }\textbf {\bibinfo
  {volume} {422}},\ \bibinfo {pages} {415} (\bibinfo {year}
  {2003})}\BibitemShut {NoStop}%
\bibitem [{\citenamefont {Cho}\ \emph {et~al.}(2015)\citenamefont {Cho},
  \citenamefont {Dellabetta}, \citenamefont {Zhong}, \citenamefont
  {Schneeloch}, \citenamefont {Liu}, \citenamefont {Gu}, \citenamefont
  {Gilbert},\ and\ \citenamefont {Mason}}]{Cho2015}%
  \BibitemOpen
  \bibfield  {author} {\bibinfo {author} {\bibfnamefont {S.}~\bibnamefont
  {Cho}}, \bibinfo {author} {\bibfnamefont {B.}~\bibnamefont {Dellabetta}},
  \bibinfo {author} {\bibfnamefont {R.}~\bibnamefont {Zhong}}, \bibinfo
  {author} {\bibfnamefont {J.}~\bibnamefont {Schneeloch}}, \bibinfo {author}
  {\bibfnamefont {T.}~\bibnamefont {Liu}}, \bibinfo {author} {\bibfnamefont
  {G.}~\bibnamefont {Gu}}, \bibinfo {author} {\bibfnamefont {M.~J.}\
  \bibnamefont {Gilbert}}, \ and\ \bibinfo {author} {\bibfnamefont
  {N.}~\bibnamefont {Mason}},\ }\href {\doibase 10.1038/ncomms8634} {\bibfield
  {journal} {\bibinfo  {journal} {Nat. Commun.}\ }\textbf {\bibinfo {volume}
  {6}},\ \bibinfo {pages} {7634} (\bibinfo {year} {2015})}\BibitemShut
  {NoStop}%
\bibitem [{\citenamefont {Maciejko}\ \emph {et~al.}(2010)\citenamefont
  {Maciejko}, \citenamefont {Kim},\ and\ \citenamefont {Qi}}]{MacIejko2010}%
  \BibitemOpen
  \bibfield  {author} {\bibinfo {author} {\bibfnamefont {J.}~\bibnamefont
  {Maciejko}}, \bibinfo {author} {\bibfnamefont {E.-A.}\ \bibnamefont {Kim}}, \
  and\ \bibinfo {author} {\bibfnamefont {X.-L.}\ \bibnamefont {Qi}},\ }\href
  {\doibase 10.1103/PhysRevB.82.195409} {\bibfield  {journal} {\bibinfo
  {journal} {Phys. Rev. B}\ }\textbf {\bibinfo {volume} {82}},\ \bibinfo
  {pages} {195409} (\bibinfo {year} {2010})}\BibitemShut {NoStop}%
\bibitem [{\citenamefont {Virtanen}\ and\ \citenamefont
  {Recher}(2011)}]{Virtanen2011}%
  \BibitemOpen
  \bibfield  {author} {\bibinfo {author} {\bibfnamefont {P.}~\bibnamefont
  {Virtanen}}\ and\ \bibinfo {author} {\bibfnamefont {P.}~\bibnamefont
  {Recher}},\ }\href {\doibase 10.1103/PhysRevB.83.115332} {\bibfield
  {journal} {\bibinfo  {journal} {Phys. Rev. B}\ }\textbf {\bibinfo {volume}
  {83}},\ \bibinfo {pages} {115332} (\bibinfo {year} {2011})}\BibitemShut
  {NoStop}%
\bibitem [{\citenamefont {Dolcini}(2011)}]{Dolcini2011}%
  \BibitemOpen
  \bibfield  {author} {\bibinfo {author} {\bibfnamefont {F.}~\bibnamefont
  {Dolcini}},\ }\href {\doibase 10.1103/PhysRevB.83.165304} {\bibfield
  {journal} {\bibinfo  {journal} {Phys. Rev. B}\ }\textbf {\bibinfo {volume}
  {83}},\ \bibinfo {pages} {165304} (\bibinfo {year} {2011})}\BibitemShut
  {NoStop}%
\bibitem [{\citenamefont {Chalker}\ and\ \citenamefont
  {Coddington}(1988)}]{Chalker1988}%
  \BibitemOpen
  \bibfield  {author} {\bibinfo {author} {\bibfnamefont {J.~T.}\ \bibnamefont
  {Chalker}}\ and\ \bibinfo {author} {\bibfnamefont {P.~D.}\ \bibnamefont
  {Coddington}},\ }\href {\doibase 10.1088/0022-3719/21/14/008} {\bibfield
  {journal} {\bibinfo  {journal} {J. Phys. C Solid State Phys.}\ }\textbf
  {\bibinfo {volume} {21}},\ \bibinfo {pages} {2665} (\bibinfo {year}
  {1988})}\BibitemShut {NoStop}%
\bibitem [{\citenamefont {Po}\ \emph {et~al.}(2018)\citenamefont {Po},
  \citenamefont {Zou}, \citenamefont {Vishwanath},\ and\ \citenamefont
  {Senthil}}]{Po2018}%
  \BibitemOpen
  \bibfield  {author} {\bibinfo {author} {\bibfnamefont {H.~C.}\ \bibnamefont
  {Po}}, \bibinfo {author} {\bibfnamefont {L.}~\bibnamefont {Zou}}, \bibinfo
  {author} {\bibfnamefont {A.}~\bibnamefont {Vishwanath}}, \ and\ \bibinfo
  {author} {\bibfnamefont {T.}~\bibnamefont {Senthil}},\ }\href {\doibase
  10.1103/PhysRevX.8.031089} {\bibfield  {journal} {\bibinfo  {journal} {Phys.
  Rev. X}\ }\textbf {\bibinfo {volume} {8}},\ \bibinfo {pages} {031089}
  (\bibinfo {year} {2018})}\BibitemShut {NoStop}%
\bibitem [{\citenamefont {Zou}\ \emph {et~al.}(2018)\citenamefont {Zou},
  \citenamefont {Po}, \citenamefont {Vishwanath},\ and\ \citenamefont
  {Senthil}}]{Zou2018}%
  \BibitemOpen
  \bibfield  {author} {\bibinfo {author} {\bibfnamefont {L.}~\bibnamefont
  {Zou}}, \bibinfo {author} {\bibfnamefont {H.~C.}\ \bibnamefont {Po}},
  \bibinfo {author} {\bibfnamefont {A.}~\bibnamefont {Vishwanath}}, \ and\
  \bibinfo {author} {\bibfnamefont {T.}~\bibnamefont {Senthil}},\ }\href
  {\doibase 10.1103/PhysRevB.98.085435} {\bibfield  {journal} {\bibinfo
  {journal} {Phys. Rev. B}\ }\textbf {\bibinfo {volume} {98}},\ \bibinfo
  {pages} {085435} (\bibinfo {year} {2018})}\BibitemShut {NoStop}%
\bibitem [{Note1()}]{Note1}%
  \BibitemOpen
  \bibinfo {note} {See supplemental material [url to be added].}\BibitemShut
  {Stop}%
\bibitem [{\citenamefont {Qiao}\ \emph {et~al.}(2014)\citenamefont {Qiao},
  \citenamefont {Jung}, \citenamefont {Lin}, \citenamefont {Ren}, \citenamefont
  {MacDonald},\ and\ \citenamefont {Niu}}]{Qiao2014a}%
  \BibitemOpen
  \bibfield  {author} {\bibinfo {author} {\bibfnamefont {Z.}~\bibnamefont
  {Qiao}}, \bibinfo {author} {\bibfnamefont {J.}~\bibnamefont {Jung}}, \bibinfo
  {author} {\bibfnamefont {C.}~\bibnamefont {Lin}}, \bibinfo {author}
  {\bibfnamefont {Y.}~\bibnamefont {Ren}}, \bibinfo {author} {\bibfnamefont
  {A.~H.}\ \bibnamefont {MacDonald}}, \ and\ \bibinfo {author} {\bibfnamefont
  {Q.}~\bibnamefont {Niu}},\ }\href {\doibase 10.1103/PhysRevLett.112.206601}
  {\bibfield  {journal} {\bibinfo  {journal} {Phys. Rev. Lett.}\ }\textbf
  {\bibinfo {volume} {112}},\ \bibinfo {pages} {206601} (\bibinfo {year}
  {2014})}\BibitemShut {NoStop}%
\bibitem [{\citenamefont {Pal}\ \emph {et~al.}(2019)\citenamefont {Pal},
  \citenamefont {Spitz},\ and\ \citenamefont {Kindermann}}]{Pal2019}%
  \BibitemOpen
  \bibfield  {author} {\bibinfo {author} {\bibfnamefont {H.~K.}\ \bibnamefont
  {Pal}}, \bibinfo {author} {\bibfnamefont {S.}~\bibnamefont {Spitz}}, \ and\
  \bibinfo {author} {\bibfnamefont {M.}~\bibnamefont {Kindermann}},\ }\href
  {\doibase 10.1103/PhysRevLett.123.186402} {\bibfield  {journal} {\bibinfo
  {journal} {Phys. Rev. Lett.}\ }\textbf {\bibinfo {volume} {123}},\ \bibinfo
  {pages} {186402} (\bibinfo {year} {2019})}\BibitemShut {NoStop}%
\end{thebibliography}%

%%%%%%%%%%%%%%%%%
% SUPPLEMENTAL MATERIAL %
%%%%%%%%%%%%%%%%%

\pagebreak
\blankpage
\onecolumngrid

\begin{center}
\textbf{\large Supplemental Material for ``Aharonov-Bohm Oscillations in Minimally Twisted Bilayer Graphene"}
\end{center}

% reset counters
\setcounter{equation}{0}
\setcounter{figure}{0}
\setcounter{table}{0}
\setcounter{page}{1}

% number supplemental material with 'S'
\renewcommand{\thepage}{S\arabic{page}}
\renewcommand{\thesection}{S\arabic{section}}  
\renewcommand{\thetable}{S\arabic{table}}  
\renewcommand{\thefigure}{S\arabic{figure}}

\section{S1.\quad Symmetry constraints on the S matrix}

First, we consider $C_3$ rotation symmetry which preserves the valley:
\begin{equation}
\mathcal S = C_3 \mathcal S C_3^{-1},
\end{equation}
where $C_3$ is a cyclic permutation $(a_1,a_1') \rightarrow (a_2,a_2') \rightarrow (a_3,a_3') \rightarrow (a_1,a_1')$ of the incoming modes which are defined in Fig.\ {\color{red} 1}(b) of the main text, and similar for outgoing modes. Next, we discuss the effect of $C_2$ rotation symmetry and time-reversal symmetry $T$. As these symmetries do not conserve the valley, we need to consider both valleys:
\begin{equation}
\begin{pmatrix} b_K \\ b_{K'} \end{pmatrix} = \begin{pmatrix} \mathcal S_K & 0 \\ 0 & \mathcal S_{K'} \end{pmatrix} \begin{pmatrix} a_K \\ a_{K'} \end{pmatrix}.
\end{equation}
In the basis shown in Fig.\ {\color{red} 1}(b) of the main text, $C_2$ rotation symmetry gives
\begin{equation}
\begin{pmatrix} b_{K'} \\ b_K \end{pmatrix} = \begin{pmatrix} \mathcal S_K & 0 \\ 0 & \mathcal S_{K'} \end{pmatrix} \begin{pmatrix} a_{K'} \\ a_K \end{pmatrix},
\end{equation}
such that $\mathcal S_{K'} = \mathcal S_K$. On the other hand, under time-reversal symmetry we have
\begin{equation}
\begin{pmatrix} a_{K'}^* \\ a_K^* \end{pmatrix} = \begin{pmatrix} \mathcal S_K & 0 \\ 0 & \mathcal S_{K'} \end{pmatrix} \begin{pmatrix} b_{K'}^* \\ b_K^* \end{pmatrix},
\end{equation}
such that $\mathcal S_{K'} = (\mathcal S_K)^t$. Hence, the combination $C_2T$ enforces $\mathcal S_K = (\mathcal S_K)^t$.

\section{S2.\quad S matrix without forward scattering}

\begin{figure}
\centering
\includegraphics[width=\linewidth]{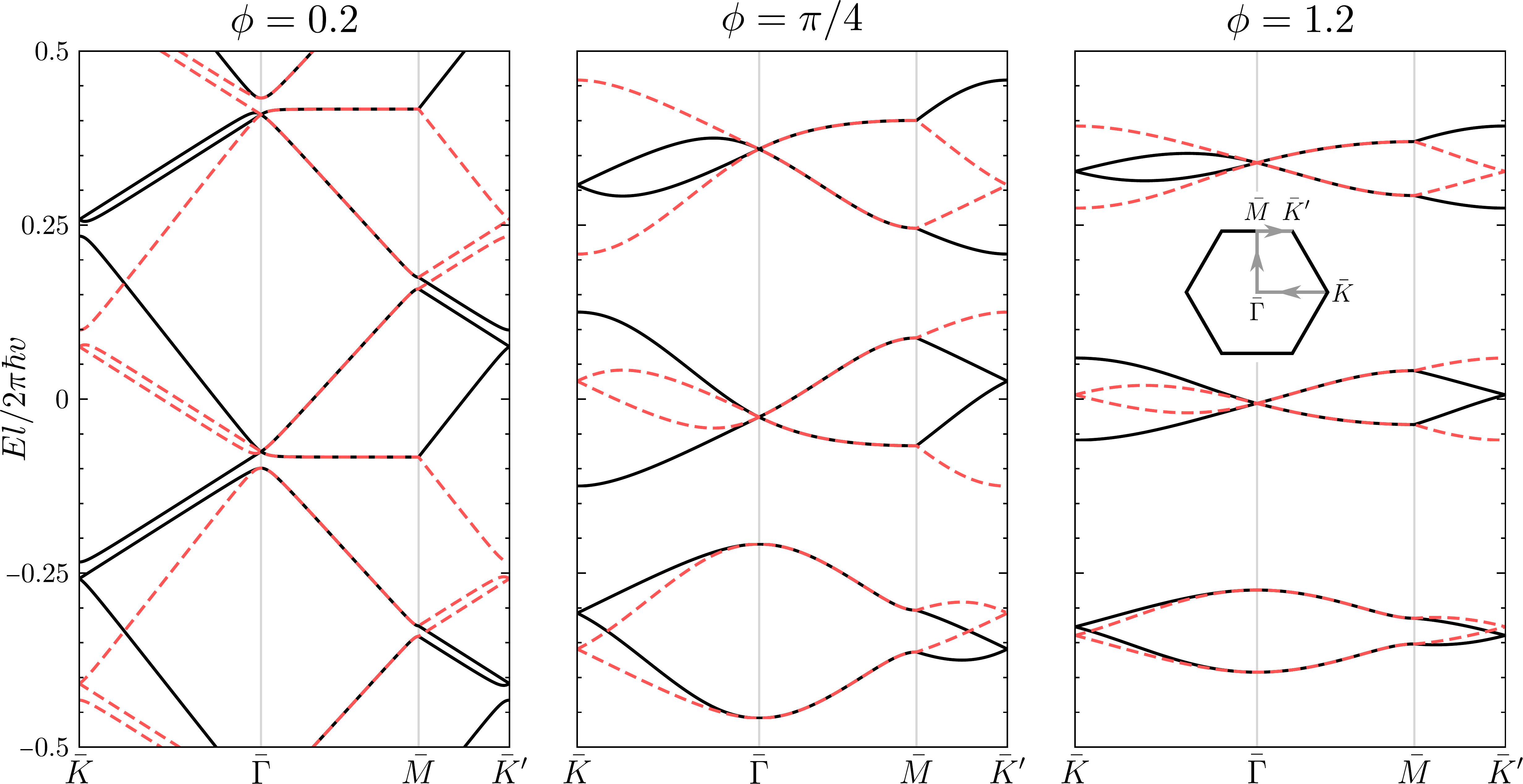}
\caption{Network spectrum in the absence of forward scattering over one energy period $2\pi \hbar v/l$ along high-symmetry lines of the moir\'e Brillouin zone, as shown in the inset of the rightmost panel. Here, we put $\varphi=-\pi \hbar v/6l$ which centers the flatbands for $\phi=\pi/2$ around zero energy,  and solid (dashed) curves correspond to the $K$ ($K'$) valley.}
\label{fig:sfig1}
\end{figure}

The $S$ matrix relates valley Hall states that propagate along AB/BA domain walls at the scattering nodes (AA regions) such that $(b,b')^t = \mathcal S (a, a')^t$ with $a,a'$ six incoming modes and $b,b'$ six outgoing modes where the prime distinguishes the two valley Hall states as illustrated in Fig.\ {\color{red} 1}(b) of the main text. In the absence of forward scattering, we find that the most general $S$ matrix consistent with unitarity and $C_3$ and $C_2T$ symmetry is given by
\begin{equation} \label{eq:Smat1}
\mathcal S(\phi,P_{d1}) = e^{i\varphi} \begin{pmatrix}
0 & e^{i \phi} \sqrt{P_{d1}} & e^{i \phi} \sqrt{P_{d1}} & 0 & \sqrt{P_{d2R}} & -\sqrt{P_{d2L}} \\
e^{i \phi} \sqrt{P_{d1}} & 0 & e^{i \phi} \sqrt{P_{d1}} & -\sqrt{P_{d2L}} & 0 & \sqrt{P_{d2R}} \\
e^{i \phi} \sqrt{P_{d1}} & e^{i \phi} \sqrt{P_{d1}} & 0 & \sqrt{P_{d2R}} & -\sqrt{P_{d2L}} & 0 \\
0 & -\sqrt{P_{d2L}} & \sqrt{P_{d2R}} & 0 & -e^{-i\phi} \sqrt{P_{d1}} & -e^{-i\phi} \sqrt{P_{d1}} \\
\sqrt{P_{d2R}} & 0 & -\sqrt{P_{d2L}} & -e^{-i\phi} \sqrt{P_{d1}} & 0 & -e^{-i\phi} \sqrt{P_{d1}} \\
-\sqrt{P_{d2L}} & \sqrt{P_{d2R}} & 0 & -e^{-i\phi} \sqrt{P_{d1}} &-e^{-i\phi} \sqrt{P_{d1}} & 0
\end{pmatrix},
\end{equation}
with $\varphi$ and $\phi$ real, $P_{d1}$ the intrachannel deflection probability, and $P_{d2R}$ ($P_{d2L}$) the probability for interchannel deflections to the right (left). Unitarity gives the conditions $2P_{d1}+P_{d2R}+P_{d2L}=1$ and $P_{d1}=\sqrt{P_{d2R} P_{d2L}}$. This has two solutions: either all probabilities are nonzero with $0<P_{d1} \leq 1/4$ the only independent parameter or $P_{d1}=0$ and either $P_{d2R}$ or $P_{d2L}$ zero, which is equivalent to what we call the zigzag regime below. Hence, we consider the former solution. In this case, the secular equations yields
\begin{equation} \label{eq:secular1}
1 - \lambda^6 + \lambda^2 \left[ \lambda^2 h(\bm k) - h(\bm k)^* \right] \left( 1 - 4 P_{d1} \sin^2 \phi \right) + 2i \lambda^3 ( 2 \sqrt{P_{d1}} \sin \phi )^3=0,
\end{equation}
with $\lambda = e^{i(El/\hbar v+\varphi)}$ and $h(\bm k)=e^{ik_1}+e^{ik_2}+e^{-i(k_1+k_2)}$. If we define $\sin \phi' = 2\sqrt{P_{d1}} \sin \phi$, which always has a solution for $\phi'$ since $0<P_{d1} \leq 1/4$, we can write the secular equation as 
\begin{equation} \label{eq:secular2}
1 - \lambda^6 + \lambda^2 \left[ \lambda^2 h(\bm k) - h(\bm k)^* \right] \cos^2 \phi' + 2i \lambda^3 \sin^3 \phi'=0,
\end{equation}
which is equivalent to the case $P_{d1}=P_{d2R}=P_{d2L}=1/4$ and $\phi \rightarrow \phi'$. Hence, it is reasonable to assume that $\mathcal S(\phi,P_{d1})$ is unitary equivalent to $\mathcal S(\phi',1/4)$. The latter $S$ matrix is given in Eq.\ ({\color{red} 1}) of the main text with $\varphi=0$ as it only shifts the overal energy, and where we drop the prime on $\phi$ from now on. For general $\phi$, Eq.\ \eqref{eq:secular2} has analytical solutions only at $\bar \Gamma$, in which case $h=3$ and we find
\begin{equation}
\lambda_{1\pm} = \pm e^{\mp i \phi}, \qquad \lambda_{2\pm} = \pm \exp \left[ \pm i\arctan \left( \frac{\sin \phi}{\sqrt{4-\sin^2\phi}} \right) \right],
\end{equation}
where the $\lambda_{2\pm}$ are doubly degenerate. Hence, for each network energy period $2\pi \hbar v/l$, there are always two protected nodes at the $\bar \Gamma$ point. This is shown in Fig.\ \ref{fig:sfig1} where we show the network spectrum along high-symmetry lines of the moir\'e Brillouin zone (MBZ) for several values of $\phi$. The same statement also holds at $\bar K$ and $\bar K'$. Furthermore, when $\phi\neq n\pi$ ($n \in \mathbb Z$), we find that the triple degeneracy at the $\bar \Gamma$, $\bar K$, and $\bar K'$ points of the MBZ is lifted, while the bands remain doubly degenerate at these points for all $\phi$, even after including forward scattering, as these crossings are protected by $C_3$ and $C_2 T$ symmetry.

We have shown that up to a unitary transformation, the most general $S$ matrix in the absence of forward scattering is given by $\mathcal S(\phi,1/4)$ for which the left and right interchannel deflection amplitudes are equal. Hence, we consider this case from now on and perform the unitary transformation
\begin{equation}
U \mathcal S U^\dag = e^{i\varphi} \begin{pmatrix}
i \sin \phi \, s_0 & \cos \phi \, s_0^t \\
\cos \phi \, s_0 & i \sin \phi \, s_0^t
\end{pmatrix}, \qquad \textrm{with} \qquad s_0 = \begin{pmatrix} 0 & 0 & 1 \\ 1 & 0 & 0 \\ 0 & 1 & 0 \end{pmatrix},
\end{equation}
where $U = e^{-i\pi \sigma_y/4} e^{i\phi\sigma_z/2} \otimes \mathds 1_3$ transforms $(a,a') \rightarrow (a_+,a_-)$ with $a_\pm = ( a e^{i \phi / 2} \mp a' e^{-i \phi / 2})/\sqrt{2}$ and similar for outgoing modes. We see that for $\phi=n \pi$ ($n \in \mathbb Z$),
\begin{equation}
U \mathcal S U^\dag =  (-1)^n e^{i\varphi} \begin{pmatrix} 0 & s_0^t \\ s_0 & 0 \end{pmatrix},
\end{equation}
such that scattering modes form three independent chiral zigzag channels. Furthermore, in this case we see from Eqs.\ \eqref{eq:secular1} and \eqref{eq:secular2} that $\phi'=\phi$ such that the network supports chiral zigzag modes for any allowed values of the deflection probabilities. On the other hand, for $\phi = (n+1/2) \pi$ ($n \in \mathbb Z$),
\begin{equation}
U \mathcal S U^\dag = (-1)^n i e^{i\varphi} \begin{pmatrix} s_0 & 0 \\ 0 & s_0^t \end{pmatrix},
\end{equation}
such that scattering modes perform closed orbits around AB and BA domains.

\section{S3.\quad S matrix with forward scattering}

\begin{figure}
\centering
\includegraphics[width=\linewidth]{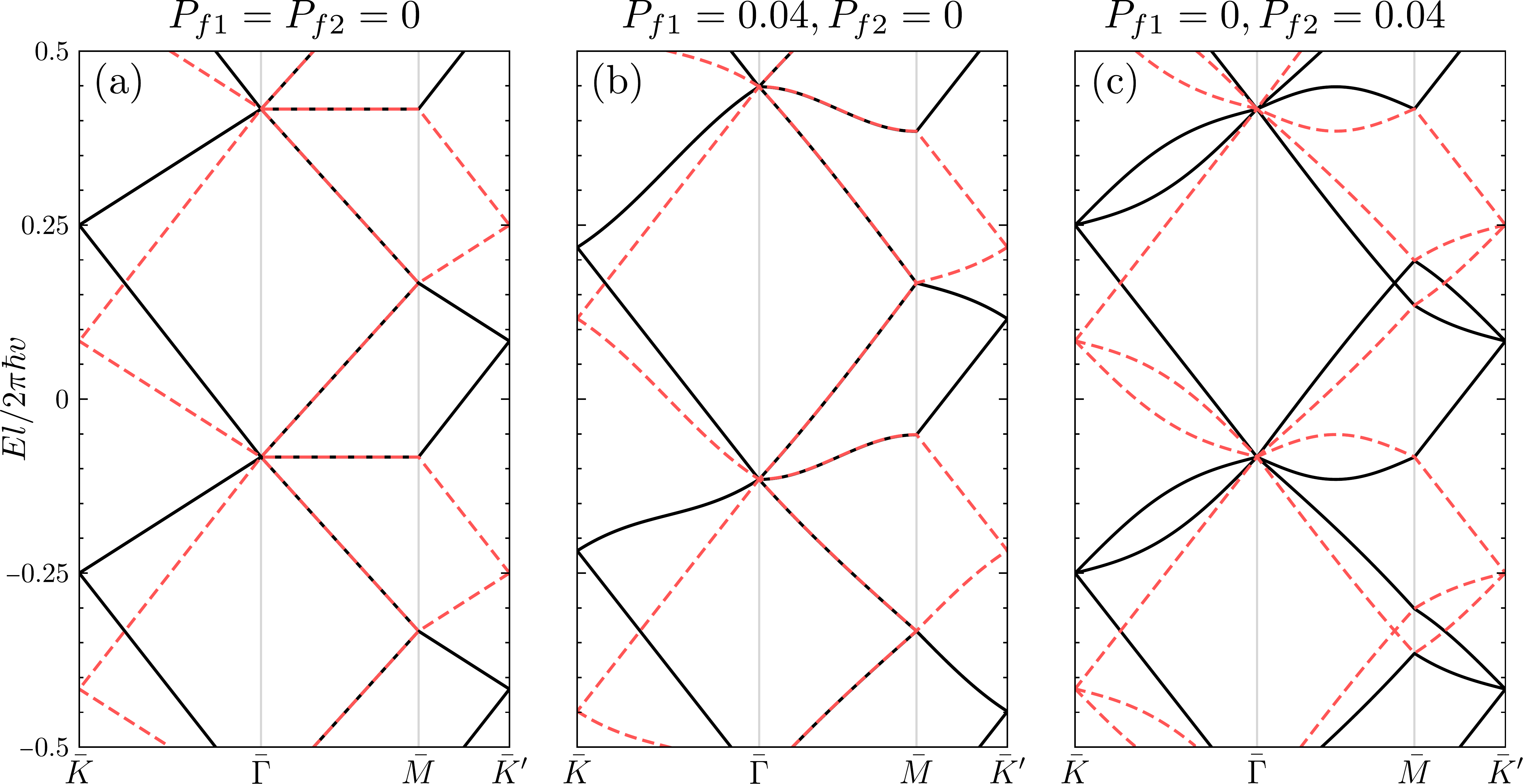}
\caption{Network spectrum for the case where zigzag modes along different directions are decoupled ($\phi=0$ and $P_{d1}=P_{d2}$) over one energy period $2\pi \hbar v/l$ along high-symmetry lines of the moir\'e Brillouin zone with $\varphi =-\pi \hbar v/6l$. The bands are shown for $K$ (solid) and $K'$ (dashed) for (a) no forward scattering, (b) $P_{f1}=0.02$ and $P_{f2}=0$, and (c) $P_{f1}=0$ and $P_{f2}=0.02$.}
\label{fig:sfig2}
\end{figure}

When we allow for forward scattering, the $S$ matrix can be written as
\begin{equation} \label{eq:Smat2}
\mathcal S =
e^{i\varphi} \begin{pmatrix}
e^{i(\phi+\chi)} \sqrt{P_{f1}} & e^{i\phi} \sqrt{P_{d1}} & e^{i\phi} \sqrt{P_{d1}} & -\sqrt{P_{f2}} & \sqrt{P_{d2}} & - \sqrt{P_{d2}} \\
e^{i\phi} \sqrt{P_{d1}} & e^{i(\phi+\chi)} \sqrt{P_{f1}} & e^{i\phi} \sqrt{P_{d1}} & -\sqrt{P_{d2}} & -\sqrt{P_{f2}} & \sqrt{P_{d2}} \\
e^{i\phi} \sqrt{P_{d1}} & e^{i\phi} \sqrt{P_{d1}} & e^{i(\phi+\chi)} \sqrt{P_{f1}} & \sqrt{P_{d2}} & - \sqrt{P_{d2}} & -\sqrt{P_{f2}} \\
-\sqrt{P_{f2}} & -\sqrt{P_{d2}} & \sqrt{P_{d2}} & -e^{-i(\phi+\chi)} \sqrt{P_{f1}} & -e^{-i\phi} \sqrt{P_{d1}} & -e^{-i\phi} \sqrt{P_{d1}} \\
\sqrt{P_{d2}} & -\sqrt{P_{f2}} & -\sqrt{P_{d2}} & -e^{-i\phi} \sqrt{P_{d1}} & -e^{-i(\phi+\chi)} \sqrt{P_{f1}} & -e^{-i\phi} \sqrt{P_{d1}} \\
-\sqrt{P_{d2}} & \sqrt{P_{d2}} & -\sqrt{P_{f2}} & -e^{-i\phi} \sqrt{P_{d1}} & -e^{-i\phi} \sqrt{P_{d1}} & -e^{-i(\phi+\chi)} \sqrt{P_{f1}}
\end{pmatrix},
\end{equation}
with $\cos\chi = \left( P_{d2} - P_{d1} \right) / 2 \sqrt{P_{f1} P_{d1}}$ such that $\chi$ is real, so that $2 \sqrt{P_{f1} P_{d1}} \geq | P_{d2}-P_{d1} |$. Note that when $P_{d2}=0$, this condition gives a lower bound on forward scattering $P_{f1} \geq P_{d1}/2$. Here, we assumed that the probability for intrachannel processes is the same for the two valley Hall states. Current conservation then requires $2(P_{d1} + P_{d2}) + P_{f1} + P_{f2} = 1$, where $P_{f1}$ ($P_{d1}$) and $P_{f2}$ ($P_{d2}$) are the probabilities for intra- and interchannel forward scattering (deflections), respectively. To investigate the couplings between zigzag modes, it is preferable to work in the new basis: $(b_+,b_-)^t = U\mathcal SU^{-1}(a_+,a_-)^t$. For $P_d=P_{d1}=P_{d2}$, we find
\begin{equation}
U \mathcal S U^\dag = e^{i\varphi}
\begin{pmatrix}
f \cos \phi & 0 & 2i \sqrt{P_d} \sin \phi & -if^* \sin \phi & 2 \sqrt{P_d} \cos \phi & 0 \\
2i \sqrt{P_d} \sin \phi & f \cos \phi & 0 & 0 & -if^* \sin \phi & 2 \sqrt{P_d} \cos \phi \\
0 & 2i \sqrt{P_d} \sin \phi & f \cos \phi & 2 \sqrt{P_d} \cos \phi & 0 & -if^* \sin \phi \\
if \sin \phi & 0 & 2 \sqrt{P_d} \cos \phi  & -f^* \cos \phi & 2i \sqrt{P_d} \sin \phi & 0 \\
2 \sqrt{P_d} \cos \phi & if \sin \phi  & 0 & 0 & -f^* \cos \phi & 2i \sqrt{P_d} \sin \phi \\
0 & 2 \sqrt{P_d} \cos \phi & if \sin \phi  & 2i \sqrt{P_d} \sin \phi & 0 & -f^* \cos \phi
\end{pmatrix},
\end{equation}
where $f=\sqrt{P_{f2}}+i\sqrt{P_{f1}}$.

In the absence of coupling between zigzag modes propagating along different direction ($\phi=0$ and $P_{d1}=P_{d2}$), the network spectrum is given by $(j=1,2,3)$
\begin{equation} \label{eq:spectrum2}
E_{j,\pm,n}(\bm k) = \frac{\hbar v}{l} \left[ 2 \pi n - \varphi - i \log \frac{1}{2} \left( F_j(\bm k) \pm \sqrt{4e^{i k_j} + F_j(\bm k)^2} \right) \right],
\end{equation}
where $F_j(\bm k) = f^* e^{-ik_{j+1}} - f e^{-ik_{j+2}}$ with $j$ defined cyclically and which is shown in Fig.\ \ref{fig:sfig2}. As can be seen in Fig.\ {\color{red} 2}(c) of the main text, coupling between parallel zigzag channels warps the Fermi surface, in a manner that depends on the type of forward scattering. For $P_{f2}=0$, the bands are symmetric about $k_y$ as in this case $F_1(k_x,-k_y)=F_2(k_x,k_y)$ and $F_3(k_x,-k_y)=F_3(k_x,k_y)$. Moreover, states at the $\bar \Gamma$ and $\pm\bar K$ points in the moir\'e Brillouin zone (MBZ) remain triply degenerate, but are shifted as $F_j(\bar \Gamma) = -2i\sqrt{P_{f1}}$ and $F_j(\pm \bar K) = \pm 2 e^{\pm i\pi/6} \sqrt{P_{f1}}$ for all $j$.

\section{S4.\quad Transport calculation}

We consider a network strip as shown in Fig.\ \ref{fig:sfig3} with length $L = Nl$ or $(N-1/2)l$ ($N=1,2,\ldots$) and width $W \gg L$, where $l \approx 14 \left( \theta^\circ \right)^{-1} \,$nm is the moir\'e lattice constant. To investigate transport, we calculate the transfer matrix that relates amplitudes between two sides, e.g.\ $\psi_{\textrm{right}} = \mathcal M \psi_{\textrm{left}}$. For the transfer matrix, unitarity is expressed as $\mathcal M^\dag J \mathcal M = J$, with $J = \diag \left( 1, 1,-1,-1,1,1\right)$ the current operator for a bulk node, so that $|\textrm{det} \, \mathcal M| =1$.

\subsection{A.\quad Transfer-matrix method}

\begin{figure}
\centering
\includegraphics[width=.75\linewidth]{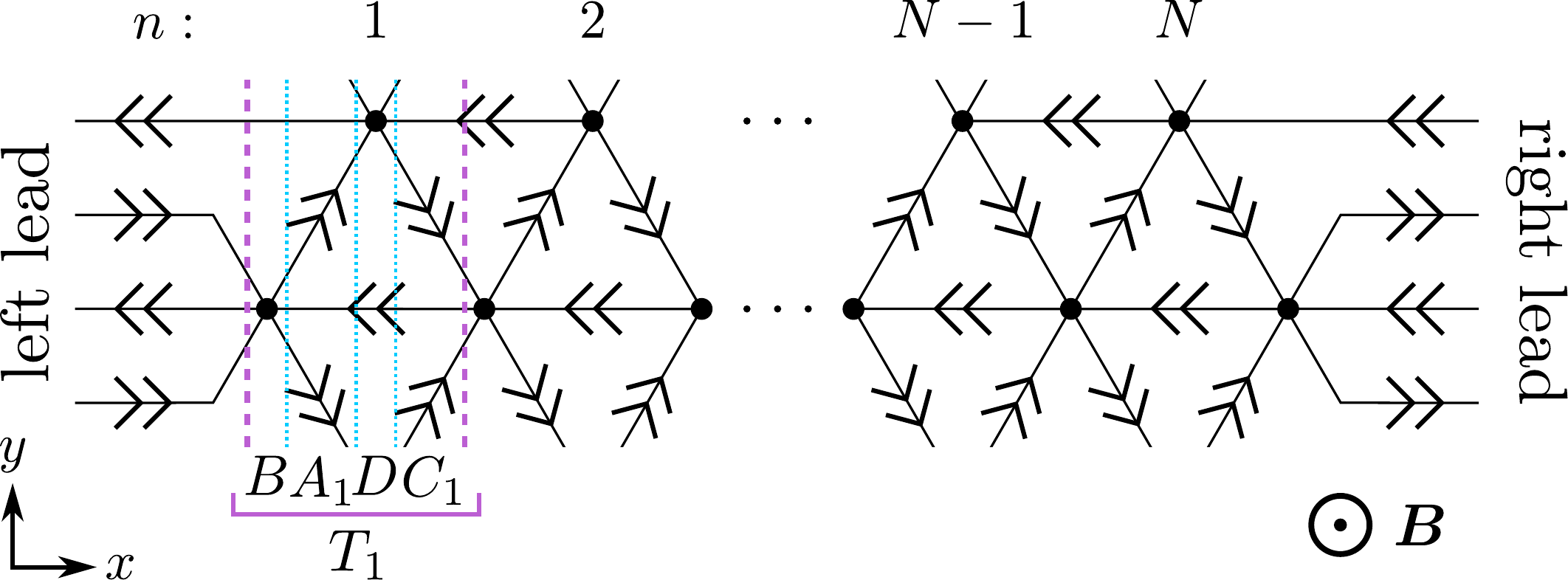}
\caption{Unit cell of a network strip with length $L=Nl$ where the dashed vertical lines outline the slice corresponding to the transfer matrix $T_1$ and dotted lines mark subslices with transfer matrices $A$, $B$, $C$, and $D$ as indicated [see Eq.\ \eqref{eq:transfer}].}
\label{fig:sfig3}
\end{figure}

The total transfer matrix $T$ links one lead to the other lead, e.g.\ $\psi_{N+1} = T \psi_0$ with
\begin{equation} \label{eq:transfer}
T = 
\begin{cases}
B T_N \cdots T_1 & \qquad L = Nl, \\
T_N \cdots T_1 & \qquad L = (N-\frac{1}{2})l,
\end{cases}
\end{equation}
where $T_n = C_n D A_n B$ with $A_n$, $B$, $C_n$, and $D$ defined in Fig.\ \ref{fig:sfig3}. We consider a network strip with $W\gg L$ and introduce the transverse momentum $0\leq k<2\pi/\sqrt{3}l$. If we label the modes from top to bottom, the transfer matrices of the unit cell can be written as $B = \diag \left(\mathds 1_2, \mathcal M \right)$ and
\begin{equation}
D(k) = 
\begin{pmatrix} \mathcal M_{22} & 0 & \mathcal M_{21} e^{ik\sqrt{3}l} \\ 0 & \mathds 1_2 & 0 \\ \mathcal M_{12} e^{-ik\sqrt{3}l} & 0 & \mathcal M_{11} \end{pmatrix}, \qquad \mathcal M = \begin{pmatrix} \mathcal M_{11} & \mathcal M_{12} \\ \mathcal M_{21} & \mathcal M_{22} \end{pmatrix},
\end{equation}
where we defined the submatrices $\mathcal M_{ij}$ of the transfer matrix $\mathcal M$ for a single node. Here, $\mathcal M_{11}$ and $\mathcal M_{22}$ are square matrices of dimension $2$ and $4$, respectively. In the absence of disorder and magnetic fields, $A_n = C_n = \diag \left( \lambda^{-1/2},  \lambda , \lambda^{-1/2}, \lambda \right) \otimes \mathds 1_2$ with $\lambda = e^{i El/\hbar v}$ the dynamical phase, independent of $n$. 

The transmission probability is then calculated by imposing boundary conditions on the leads. For transport in the positive $x$ direction, we have $\psi_0^{(\alpha)} = \left( r_{1\alpha}, a_{1\alpha} , r_{2\alpha}, a_{2\alpha} \right)^t$ and $\psi_{N+1}^{(\alpha)} = \left( 0, t_{1\alpha}, 0, t_{2\alpha} \right)^t$ where $r_i$ ($t_i$) are two-component reflection (transmission) amplitudes, and $\alpha$ labels the incoming modes with amplitudes $a_{11}=a_{23}=(1,0)$ and $a_{12}=a_{24}=(0,1)$ and $(0,0)$ otherwise. The transmission probability for a transverse mode with momentum $k$ is then given by $\mathcal T_k = 1 - (1/4) \sum_{\alpha=1}^4 r_\alpha^\dag r_\alpha$ with $r_\alpha = \left( r_{1\alpha} , r_{2\alpha} \right)^t$. The total transmission probability becomes 
\begin{equation}
\mathcal T = (\sqrt{3}l/2\pi) \int_0^{2\pi/\sqrt{3}l} dk \, \mathcal T_k.
\end{equation}

Up to now, we considered a single valley. For the other valley, the propagation direction of valley Hall states shown in Fig.\ \ref{fig:sfig3} is reversed, as the valleys are related by time reversal. Hence, the transmission for $K'$ in the positive $x$ direction is given by the transmission of $K$ in the negative $x$ direction. It follows that $\mathcal T_K = \mathcal T_{K'}$ in the limit $W \gg L$.

\subsection{B.\quad Magnetotransport}

The magnetic field introduces an additional Peierls phase accumulated by the valley Hall states during propagation along links. In the Landau gauge $\bm A = Bx \bm e_y$, the Peierls phase along a link starting at $x=ml/2$ ($m \in \mathbb Z$) is zero for horizontal links (see Fig.\ \ref{fig:sfig3}) and for an upward or downward diagonal link,
\begin{equation}
\Phi_{\pm}(m) = \mp \pi \left( m + \frac{1}{2} \right) \frac{\Phi}{\Phi_0} ,
\end{equation}
respectively, where $\Phi_0 = h/e$ is the flux quantum and $\Phi = B \mathcal A$ is the flux through the moir\'e cell, comprising an AB and BA triangle, with $B$ the magnetic field and $\mathcal A = \sqrt{3} \, l^2/2$ the moir\'e cell area. If we assume that the magnetic field is sufficiently small such that the structure of the $S$ matrix is unchanged, the transfer matrices $B$ and $D$ remain the same, while $A_n(\Phi) = Q_{2n-2}(\Phi)$ and $C_n(\Phi) = Q_{2n-1}(-\Phi)$ with $Q_m = \diag \left( \lambda^{-1/2},  \lambda e^{i\Phi_+(m)} , \lambda^{-1/2}, \lambda e^{i\Phi_-(m)} \right) \otimes \mathds 1_2$.

\subsubsection*{Quasi-1D regime}

In the quasi-1D regime ($P_d=P_{d1}=P_{d2}$ and $\phi =0$), there are three separate contributions to the conductance from zigzag (ZZ) modes along the $\bm l_{1,2,3}$ directions for a given valley and spin. In the setup shown in Fig.\ \ref{fig:sfig3}, the contribution from $\bm l_3$ is always given by $\mathcal T_{\bm l_3} = 2$, and the transmission for $\bm l_{1,2}$ is identical. The latter vanish in the absence of forward scattering and can be calculated straightforwardly for an infinitely wide system of length $L/l=1/2,1,3/2$ by summing Feynman paths. The results are
\begin{figure}
\centering
\includegraphics[width=.7\linewidth]{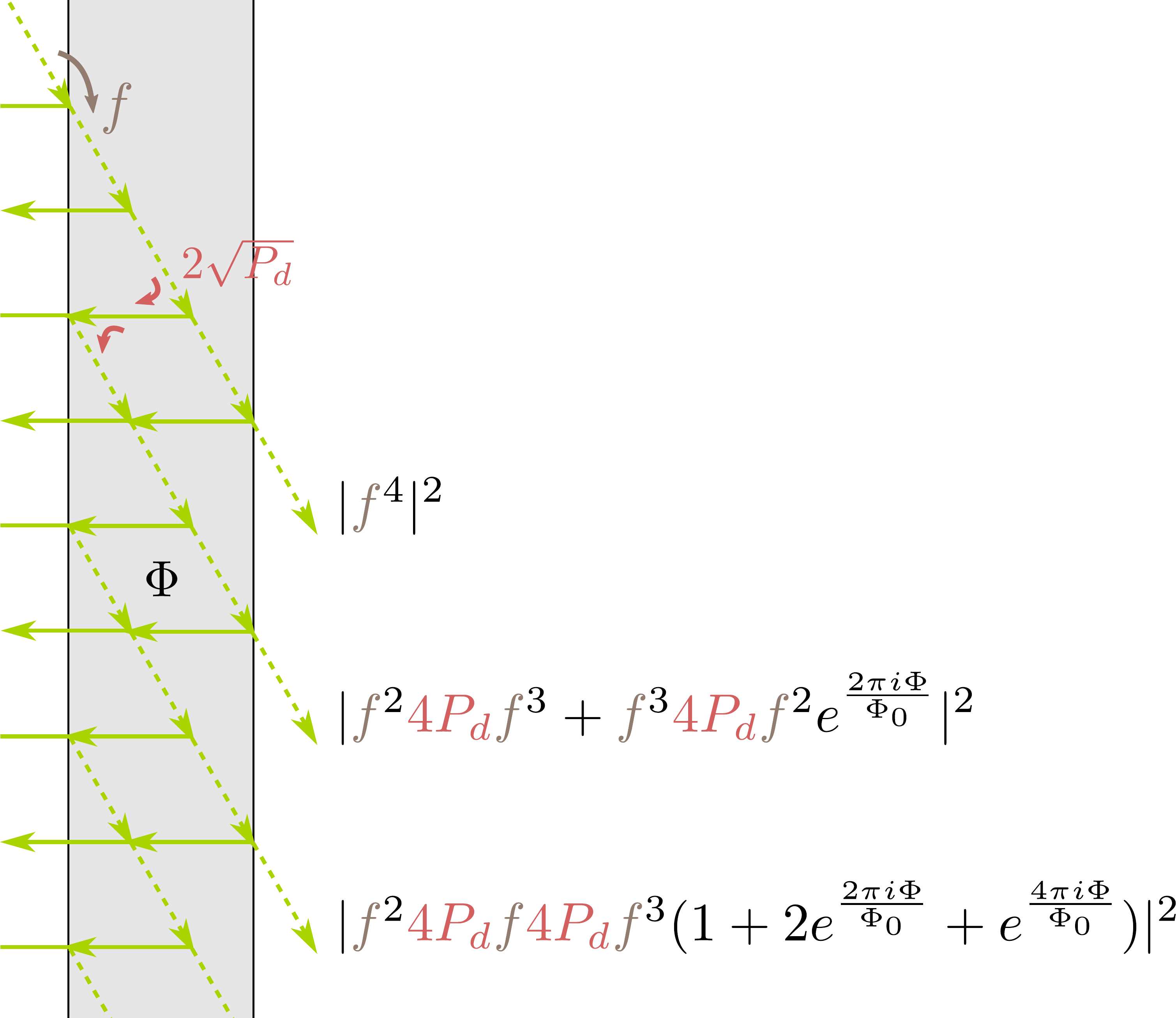}
\caption{Feynman paths in the quasi-1D regime of the $\bm l_2$ ZZ modes for a network strip of length $L=3l/2$, where $f=\sqrt{P_{f2}} + i \sqrt{P_{f1}}$ is the forward scattering amplitude in the ZZ basis between solid arrows and $\Phi$ is the flux through a moir\'e cell.}
\label{fig:sfig4}
\end{figure}
\begin{alignat}{3} \label{eq:Ga} %\frac{G}{G_0}
& \left. G \right|_{L=\frac{l}{2}} && = \left( 2 + 2 P_f^2 \right) \frac{G_0W}{\sqrt{3}l}, \\
& \left. G \right|_{L=l} && = \left( 2 + \frac{2 P_f^3}{1-P_f (4P_d)^2} \right) \frac{G_0W}{\sqrt{3}l}, \\
& \left. G \right|_{L=\frac{3l}{2}} && = \left( 2 + \frac{2 P_f^4}{1-P_f (4P_d)^2\left[ 2 \cos ( \pi \Phi/\Phi_0 ) \right]^2} \right) \frac{G_0W}{\sqrt{3}l},
\end{alignat}
where $P_f = P_{f1} + P_{f2}$, which is illustrated in Fig.\ \ref{fig:sfig4} for $L=3l/2$. Note that these expressions only depend on the total forward scattering probability $P_f$, which is always the case in the quasi-1D regime. The conductance is shown in Fig.\ \ref{fig:sfig5} at zero magnetic field as a function of $P_f$ for several lengths. For larger systems, the Feynman paths are more involved due to contributions from paths that have segments that cut horizontal along the sample, opposite to the transport direction. If we calculate $\mathcal T_{\bm l_{1,2}}$ up to second order in the path length, we find
\begin{equation} \label{eq:sGa2} 
G \simeq \left( 2 + 2 P_f^{2L/l+1} \left[ 1 + P_f (4P_d)^2 \left( \frac{\sin \left[ \pi (2L/l-1) \Phi/\Phi_0 \right]}{\sin (\pi \Phi/\Phi_0)} \right)^2 \right] \right) \frac{G_0W}{\sqrt{3}l},
\end{equation}
which includes $\mathcal T_{\bm l_3} = 2$ and contributions to $\mathcal T_{\bm l_{1,2}}$ from paths with lengths $2L$ and $2L+3l$, and corresponds to Eq.\ ({\color{red} 6}) of the main text.
\begin{figure}
\centering
\includegraphics[width=.65\linewidth]{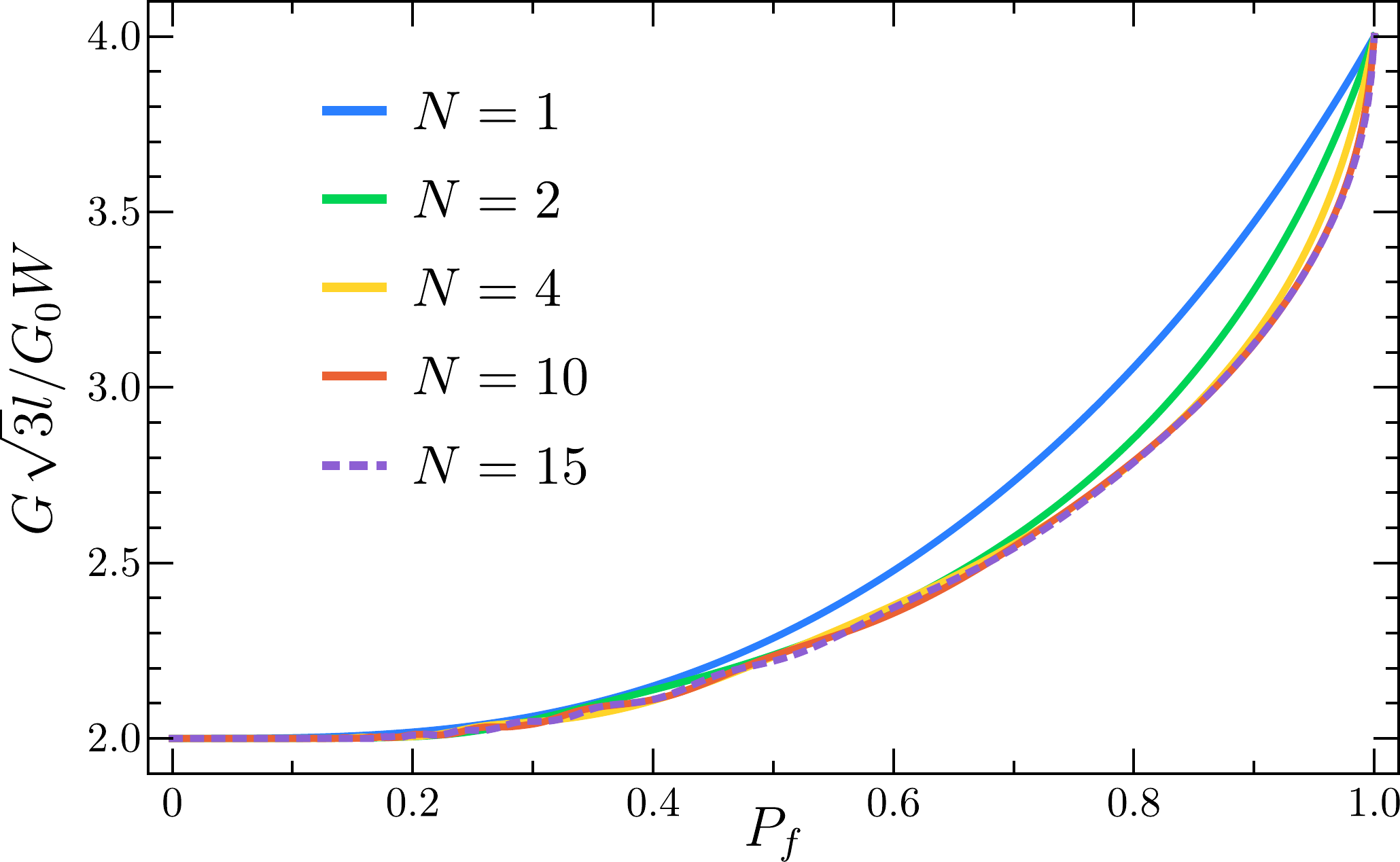}
\caption{Zero-field conductance for a network strip with width $W \gg L$ in the quasi-1D regime ($\phi=0$) as a function of the total forward scattering probability $P_f = P_{f1}+P_{f2}$ for different lengths $L = Nl$.}
\label{fig:sfig5}
\end{figure}

Remarkably, there exist formal analytical expressions for $L>3l/2$ which can be obtained by explicit calculation of the total transfer matrix and integration over transverse momentum. We find that the momentum-dependent transmission can be written as
\begin{equation}
\mathcal T_{\bm l_{1,2}}(k) = \frac{P_f^{2L/l+1}}{\sum_{n=0}^{m_L} a_n \cos^n k},
\end{equation}
where $m_L= \textrm{floor}(L/l)$ and the coefficients $a_n$ are functions of $P_f$ and the flux $\Phi$. We have observed this explicitly up to $L=2l$ for finite $\Phi$ and $L=7l/2$ for zero $\Phi$ and we expect it holds for all $L$. For example, for $L=2l$ we find that
\begin{align}
a_0 & = 1 - 4P_d P_f (1 + P_f^2) + 8 (4P_d)^2 P_f \cos^2 \frac{\pi \Phi}{\Phi_0} \cos \frac{2\pi\Phi}{\Phi_0}, \\
a_1 & = 2 \sqrt{P_f} \left[ 1 + (P_f - 2) P_f^2 \right] \left( 1 + 2 \cos \frac{2\pi\Phi}{\Phi_0} \right), \\
a_2 & = 16 P_d P_f, 
\end{align}
where $4P_d=1-P_f$. The total transmission function is given by
\begin{equation}
\mathcal T_{\bm l_{1,2}} = \frac{P_f^{2L/l+1}}{2\pi} \int_0^{2\pi} \frac{dk}{\sum_{n=0}^{m_L} a_n \cos^n k} = \frac{P_f^{2L/l+1}}{2\pi i} \oint \frac{(1/z) dz}{\sum_{n=0}^{m_L} \frac{a_n}{2^n} \left( z + 1/z \right)^n},
\end{equation}
where $z=e^{ik}$ and the contour goes along the unit circle. The integral can be worked out and becomes
\begin{equation}
\frac{1}{2\pi i} \oint \frac{z^{m_L-1} dz}{\sum_{n=0}^{2m_L} b_n z^n} = \frac{2^{m_L}}{a_{2m_L}} \left( \sum_{|z_i|<1} + \frac{1}{2} \sum_{|z_i|=1} \right) \frac{z_i^{m_L-1}}{\prod_{j\neq i} \left( z_i - z_j \right)},
\end{equation}
where $z_i$ ($i=1,\ldots,2m_L$) are the roots of the polynomial of degree $2m_L$ in the denominator on the left-hand side of the equation and we assumed that the poles are simple. If this is not the case, the results can be obtained with the general residue theorem. Moreover, up to $L=5l/2$, there exist closed-form expressions for the roots. For $L=2$, we find there are only simple poles, given by
\begin{equation}
z_{\zeta,\xi} = - \frac{a_1 + \zeta \sqrt{a_1^2-4a_0a_2} + \xi \sqrt{2a_1 \left( a_1 + \zeta \sqrt{a_1^2 - 4a_0a_2} \right) - 4a_2 (a_0 + a_2)}}{2a_2}
\end{equation}
where $\zeta,\xi=\pm$. Furthermore, we find there are always two poles inside the unit circle, labeled $z_1$ and $z_2$ (for $\Phi=0$, they are given by $z_{\pm,-}$), and two outside the unit circle, labeled $z_3$ and $z_4$, so that
\begin{equation}
\left. \mathcal T_{\bm l_{1,2}} \right|_{L=2l} = \frac{4P_f^4}{a_2} \left[ \frac{z_1}{(z_1-z_2)(z_1-z_3)(z_1-z_4)} + \frac{z_2}{(z_2-z_1)(z_2-z_3)(z_2-z_4)} \right].
\end{equation}

\subsection{C.\quad Disorder}

We distinguish three types of disorder: (1) disorder sharp on the interatomic scale that couples the valleys, (2) disorder on a larger length scale that couples the two valley Hall channels (for one valley and spin) during propagation along a link, and (3) smooth disorder that adds random phases to the dynamical phase accumulated during propagation along a link. The first type of disorder could, for example, arise from stacking faults where the twist angle changes abruptly. However, the phenomena that we want to address in this work occur in samples with an almost homogeneous twist angle. Therefore, the inclusion of stacking faults in the transport calculation is left for future research. The length scale of the second type of disorder is given by $(\hbar v /\gamma_\perp) \left( \gamma_\perp/U \right)^{1/3}$, which follows from the four-band continuum model for bilayer graphene for the case with two semi-infinite regions of AB ($x<0$) and BA ($x>0$) stacking with $\gamma_\perp \approx 0.3$~eV the interlayer coupling and $U$ the interlayer bias. This type of disorder leads to scattering between the two valley Hall states during propagation along a link. It can be included by letting
\begin{equation}
\lambda \mathds 1_6 \rightarrow \lambda \left[ \cos \tau \, \mathds 1_6 + i \sin \tau (\sigma_y \otimes \mathds 1_3) \right],
\end{equation}
where $\tau$ is the disorder strength. This couples $a_\pm$ modes along a link with probability $\cos^2 \phi \, \sin^2\tau$. Hence, it leads to couplings between ZZ modes similar to the processes shown in Fig.\ {\color{red} 2}(d) of the main text. Finally, disorder due to smooth fluctuations of the Fermi energy or the twist angle, for example, does not couple the valley Hall states along a link but introduces randomness in the scattering parameters at the nodes as well as random phases after propagation along links.

\begin{figure}
\centering
\includegraphics[width=.8\linewidth]{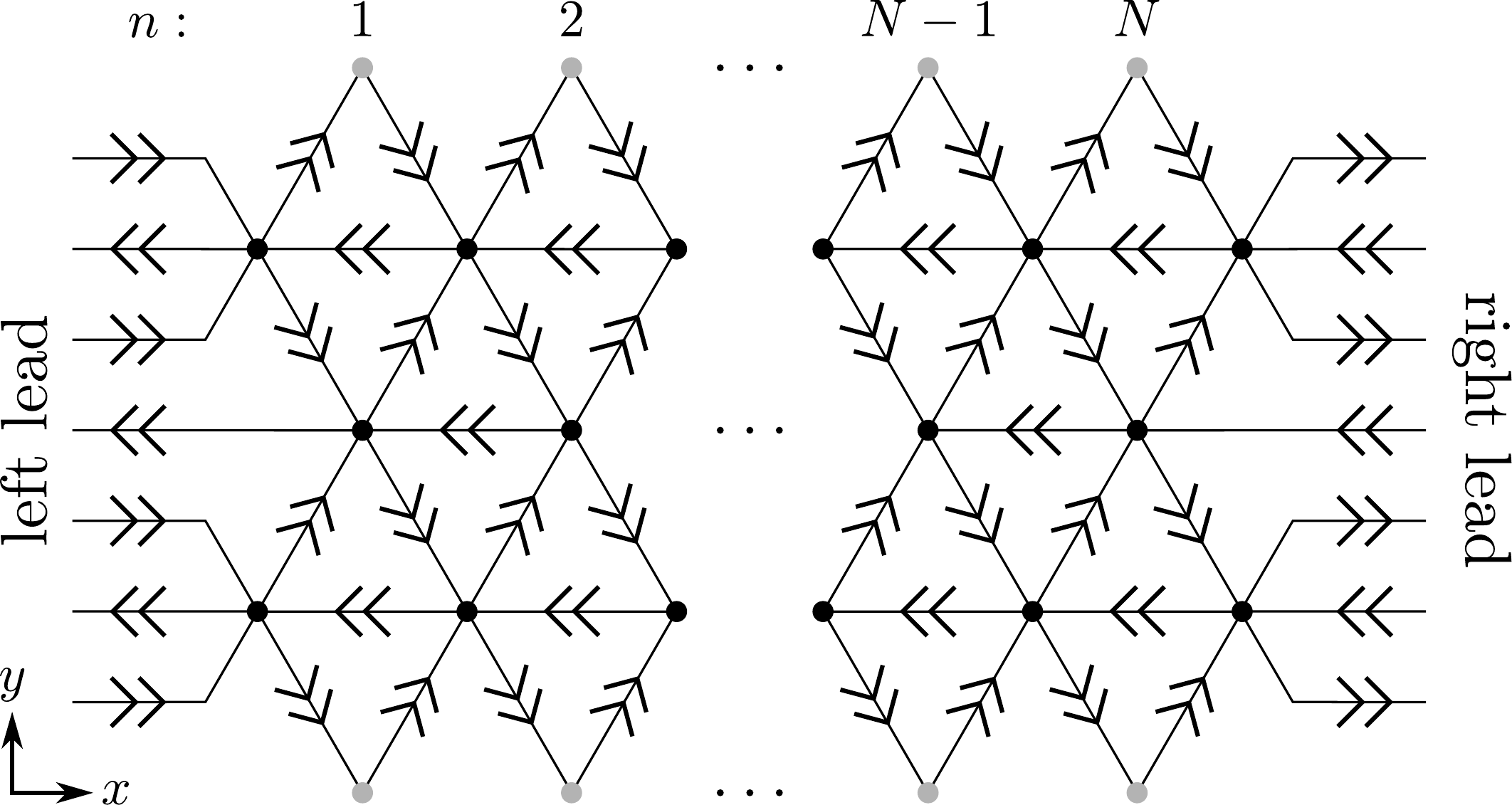}
\caption{Finite network strip of length $L=Nl$ and width $W=2\sqrt{3}l$, where black (gray) dots are bulk (edge) scattering nodes.}
\label{fig:sfig7}
\end{figure}

\begin{figure}
\centering
\includegraphics[width=\linewidth]{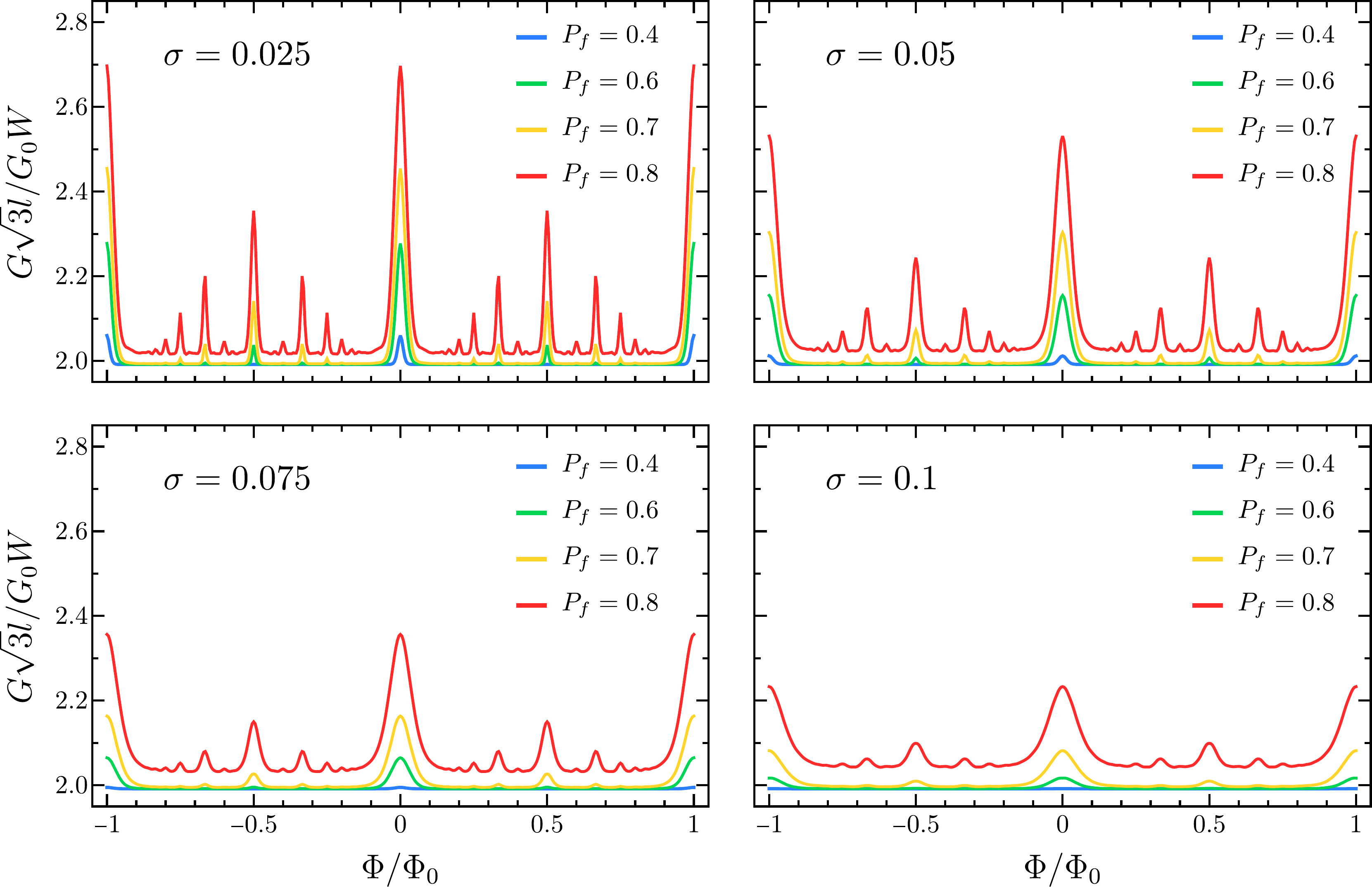}
\caption{Disorder-averaged magnetoconductance for a network strip of length $L = 10 \, l$ and width $W=60\sqrt{3} \,l$ in the quasi-1D regime ($\phi=0$) for several $P_f$ as a function of $\Phi=B\mathcal A$. The disorder strength which is given by the standard deviation $\sigma$ of the random phases is shown on the figures, and the average is over $10$ disorder configurations.}
\label{fig:sfig8}
\end{figure}

\begin{figure}
\centering
\includegraphics[width=.75\linewidth]{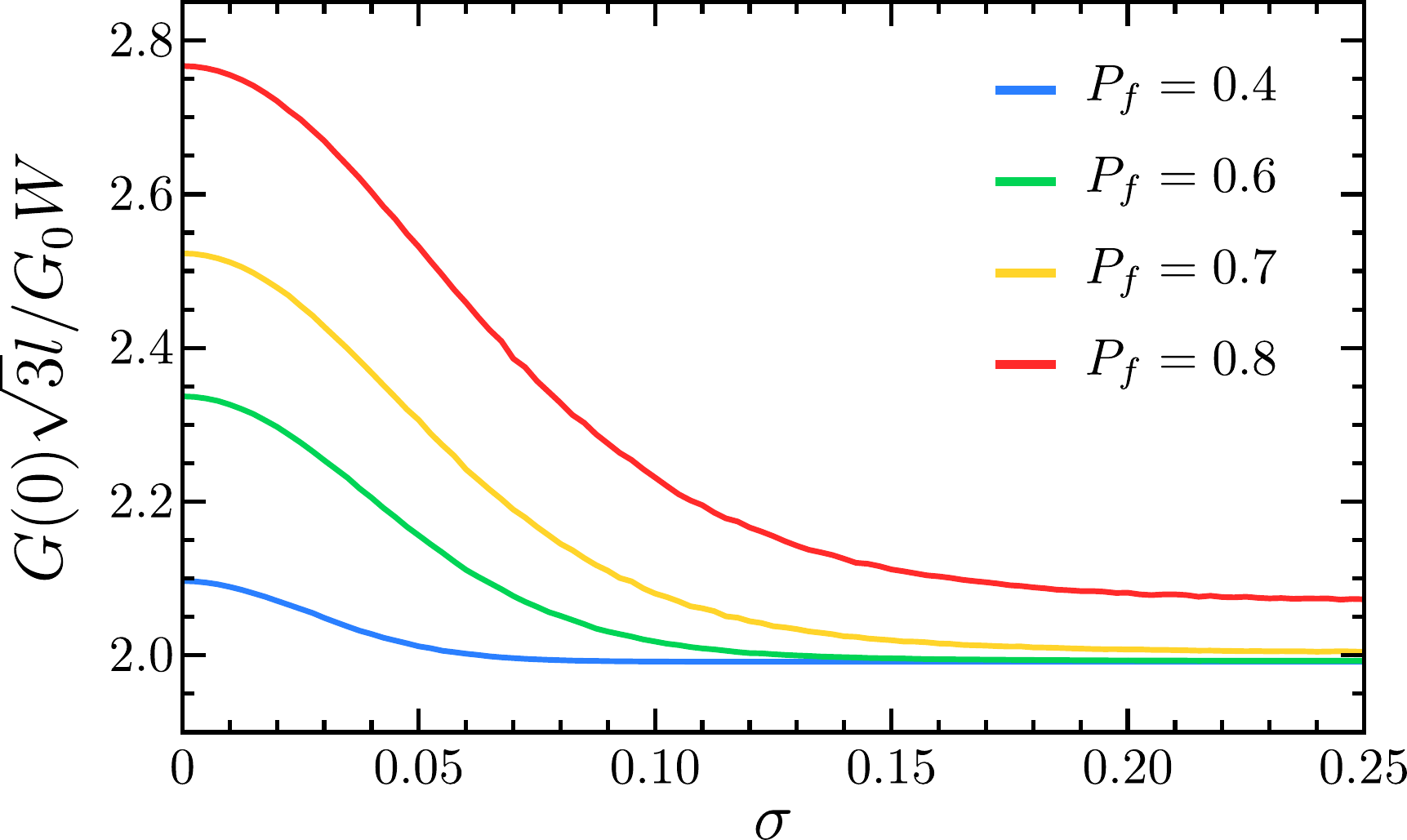}
\caption{Peak height of the conductance at $\Phi=0$ in the quasi-1D regime ($\phi=0$) as a function of the standard deviation $\sigma$ of the Gaussian-distributed random link phases with zero mean for a network strip of length $L=10\, l$ and width $W=60\sqrt{3} \,l$, averaged over $40$ disorder configurations.}
\label{fig:sfig9}
\end{figure}

We can include the third type of disorder in the transport calculations by letting $\lambda \rightarrow \lambda e^{i2\pi \gamma(\bm r)}$, where $\gamma(\bm r)$ is a normal-distributed number with zero mean and standard deviation $\sigma$, where we keep the $S$ matrix constant everywhere for simplicity. To this end, we consider a network strip with a finite width $W=\sqrt{3} Ml$, as shown in Fig.\ \ref{fig:sfig7}. We model the edges by introducing an edge $S$ matrix $\mathcal S_b$. For concreteness, we take $U\mathcal S_b U^{-1} = \sigma_x$ so that the edge does not affect $\bm l_3$ zigzag modes, but mixes the $\bm l_1$ and $\bm l_2$ zigzag modes. In any case, in the limit $W \gg L$, the boundary effect becomes negligible. In Fig.\ \ref{fig:sfig8}, we show the disorder-averaged magnetoconductance for a network strip of width $W=60\sqrt{3}\,l$ in the quasi-1D regime for the same parameters as Fig.\ {\color{red} 3}(b) of the main text with $\sigma=0.025$, $0.05$, $0.075$, and $0.1$. Note that except for the case $P_f=0.8$, the background is smaller than its value $8e^2/h$ for the infinitely-wide system, due to scattering at the edges, but which is not affected by disorder given its chiral origin. Nevertheless, we see that the A-B resonances are broadened and that the peak heights are reduced when the disorder strength $\sigma$ is increased. This can be understood from the fact that different contributions with the same path length are individually shifted in $\Phi$ by the disorder. In Fig.\ \ref{fig:sfig9}, we show the height of the main A-B peak versus $\sigma$. 

\end{document}